\documentclass[11pt,oneside,onecolumn,draftcls]{IEEEtran}
\usepackage[dvips]{graphicx}
\usepackage{epsfig}
\usepackage{amsmath}
\usepackage{epsfig}
\usepackage{array}
\usepackage{amssymb}
\usepackage{color}

\begin{document}

\title{Continuous Encryption Functions for Security Over Networks}

\author{Yingbo Hua\thanks{The authors are with
Department of Electrical and Computer Engineering,
University of California, Riverside, CA 92521, USA. Emails: yhua@ece.ucr.edu and ahmed.maksud@email.ucr.edu. This work was supported in part by the Army Research Office under Grant Number W911NF-17-1-0581 and the Department of Defense under W911NF2020267.
The views and conclusions contained in this
document are those of the authors and should not be interpreted as representing the official policies, either
expressed or implied, of the Army Research Office or the U.S. Government. The U.S. Government is
authorized to reproduce and distribute reprints for Government purposes notwithstanding any copyright
notation herein.}, \emph{Fellow, IEEE},
Ahmed Maksud, \emph{Student Member, IEEE}}

\maketitle
\begin{abstract}
This paper presents a study of continuous encryption functions (CEFs) of secret feature vectors for security over networks such as physical layer encryption for wireless communications and biometric template security for online Internet applications. CEFs are defined to include all prior continuous ``one-way'' functions. It is shown that dynamic random projection and index-of-max (IoM) hashing algorithm 1 are not hard to attack, IoM algorithm 2 is not as hard to attack as it was thought to be, and higher-order polynomials are easy to attack via substitution. Also presented is a new family of CEFs based on selected components of singular value decomposition (SVD) of a randomly modulated matrix of feature vector. Detailed empirical evidence suggests that SVD-CEF is hard to attack. Statistical analysis of SVD-CEF reveals its useful properties including its sensitivity to noise. The bit-error-rate performance of a quantized SVD-CEF is shown to exceed that of IoM algorithm 2.
\end{abstract}
%

\section{Introduction}

 Encryption is fundamentally important for information security over networks. For a vast range of situations, the amount of user's data far exceeds the amount of secrecy that is available to keep the users' data in complete secrecy. For such a situation, an often called one-way function is required to provide computation based security on top of any given amount of information-theoretic security.
 The conventional one-way functions are discrete, which in general require a secret key that is 100\% reliable.

In this paper, we are interested in applications where a reliable secret key is either not available or insufficient but a limited amount of secrecy is available in some noisy form. One such application is when
 two separated nodes (Alice and Bob) in a network do not share a secret key but they have their respective estimates of a common physical feature vector (such as reciprocal channel state information). How to use the estimated feature vectors at Alice and Bob to protect a large amount of information transmitted between them is a physical layer encryption problem  discussed in \cite{Hua2020a}-\cite{Hua2020b}. Another application is  biometric template security for Internet applications \cite{Jain2008}-\cite{Patel2015} where network users just want to rely on their own biometric feature vectors for secure online transactions.

The estimated (or measured) feature vectors are noisy. To exploit them for encryption, there are two basic approaches. The first is such that Alice and Bob attempt to generate a secret key from their noisy estimates. If successfully, this key can be then used to encrypt and decrypt a large amount of information based on a discrete encryption method. But due to noise in the estimated feature vectors, there is no guarantee that the key produced by Alice 100\% agrees with the key produced by Bob \cite{Lai2011}, \cite{Wallace2010} and \cite{Ye2010}. Any mismatched keys would generally fail a discrete encryption method. Note that an encrypted sequence is typically based on a pseudorandom sequence governed by a seed, i.e., a secret key, and a totally different pseudorandom sequence would be generated with any bit change in the seed.

The second approach is what we call here continuous encryption. For physical layer encryption \cite{Hua2020a}-\cite{Hua2020b}, for example, a message to be sent by Alice can be encrypted by a continuous encryption function (CEF) based on Alice's estimate of a secret feature vector, and the message can be then recovered by Bob using the same CEF but based on Bob's estimate of the secret feature vector.   The noises in the estimated feature vectors degrade Bob's recovery of the message but only in a soft or controllable way. This second approach  is similar in a spirit to many of the methods for  biometric template security \cite{Jain2008}-\cite{Patel2015}.

The contributions of this paper focus on a development of continuous encryption functions (CEFs). We define a CEF as any function $\mathbf{y}=f(\mathbf{x})$ that maps an $N\times 1$ \emph{continuous} real-valued vector $\mathbf{x}$ onto another $M\times 1$ real-valued vector $\mathbf{y}$. But not all CEFs have the same quality for applications.
We will only consider a CEF $\mathbf{y}=f(\mathbf{x})$ that allows $N$ and $M$ to be any positive integers and can be decomposed into $\mathbf{y}_i=f_i(\mathbf{x})$ with  $\mathbf{y}_i$ being the $i$th subvector of $\mathbf{y}$. The index $i$ here may also represent the time when the CEF is used.

If $\mathbf{y}$ is a continuous vector, we call $\mathbf{y}=f(\mathbf{x})$ a CEF of type A. If $\mathbf{y}$ is discrete, we call $\mathbf{y}=f(\mathbf{x})$ a CEF of type B. A quantization of the output of a type-A CEF converts it to a type-B CEF.

We propose to measure the (primary) quality of a CEF $\mathbf{y}=f(\mathbf{x})$ by the following criteria:
\begin{enumerate}
  \item (Hardness to invert) Can $\mathbf{x}$ be computed from $\mathbf{y}$ with a complexity order that is a polynomial function of the dimensions of $\mathbf{x}$ and $\mathbf{y}$? If yes, the function is said to be easy, or not hard, to invert. If no, the next question is important:
  \item (Hardness to substitute) Can $\mathbf{y}_i$ for some values of $i$ be written as  $\mathbf{y}_i=f_{1,i}(\mathbf{s})$ where $f_{1,i}(\mathbf{s})$ is not a hard-to-invert function of $\mathbf{s}$ and $\mathbf{s}=f_2(\mathbf{x})$ is an $i$-independent function of $\mathbf{x}$ (and has a dimension no larger than a polynomial function of $N$)? If yes, the function $f(\mathbf{x})$ is said to be easy, or not hard, to substitute. We call $\mathbf{s}=f_2(\mathbf{x})$ a substitute input.
  \item (Noise sensitivity) If the CEF is hard to invert and hard to substitute, then it is important to know how sensitive statistically $\mathbf{y}$ is to a small perturbation/noise in $\mathbf{x}$.
\end{enumerate}

The family of CEFs includes all prior hard-to-invert or one-way continuous functions proposed in the literature. The hard to invert property is widely desired in applications.  The hard to substitute property is also important for a similar reason. If an attacker is able to determine a substitute input  from prior exposed $\mathbf{y}_i$, then all future $\mathbf{y}_i$ can be predicted by the attacker.
It is clear that ``easy to invert'' implies ``easy to substitute'', but the reverse is not true in general. We say that a CEF is easy to attack if it is easy to invert \emph{or} easy to substitute, or equivalently a CEF is hard to attack if it is hard to invert \emph{and} hard to substitute.

The noise sensitivity of a CEF is also important in applications. To have a small noise sensitivity, a CEF $\mathbf{y}=f(\mathbf{x})$ must be locally continuous with probability one subject to some continuous randomness of  $\mathbf{x}$. More precisely, for a type-A CEF, we can measure its sensitivity by the square-rooted ratio of $\texttt{SNR}_\mathbf{x}$ over $\texttt{SNR}_\mathbf{y}$ where  $\texttt{SNR}_\mathbf{x}$ and $\texttt{SNR}_\mathbf{y}$ are some signal-to-noise ratios (SNRs) of $\mathbf{x}$ and $\mathbf{y}$ respectively, e.g., see \eqref{eq:eta_k_x} later. For a type-B CEF,  the sensitivity can be measured by the bit error rate (BER) in $\mathbf{y}$ caused by random perturbations in $\mathbf{x}$, which will be discussed in detail in section \ref{sec:comparison}.

A best CEF must be hard to invert and hard to substitute and have the least noise sensitivity.  It is important to note that for a given CEF, one can try to be successful to prove that the CEF is not hard to attack. But it seems impossible to \emph{prove} that a CEF is hard to attack. Such an open problem also applies to discrete one-way functions \cite{Levin2003}, \cite{KatzLindell2015}, \cite{wiki} even though their use in practice has been indispensable. We will say that a CEF is empirically hard to attack if we can provide strong empirical evidence.

\subsection{Prior Works and Current Contributions}

It appears that the prior CEFs all exploit (or can all exploit) any available secret key $S$ (as the seed) to produce pseudo-random numbers or operations needed in the functions.
The best known method to invert such a CEF in general seems to have a complexity order equal to $C_{N,M}2^{N_S}$, where $N_S$ is the number of binary bits in the secret key, and $C_{N,M}$ is the (best known) complexity to invert the CEF if the secret key is exposed. Unless mentioned otherwise, we will refer to $C_{N,M}$ as the complexity of attack. The understanding of $C_{N,M}$ is important for situations where $N_S$ is not sufficiently large or simply zero.


The random projection (RP) method in \cite{Teoh2007} and the dynamic random projection (DRP) method in \cite{Yang2010} are type-A CEFs before a quantization is applied at the last step of the functions. The Index-of-Maximum (IoM) hashing in \cite{Jin2018} is inherently a type-B CEF. The higher-order polynomials (HOP) in \cite{Grigoriev2012} are a type-A CEF.

We will show that for the RP method, the DRP method and the IoM algorithm 1, $C_{N,M}=P_{N,M}$ with $P_{N,M}$ denoting a  polynomial function of both $N$ and $M$. We will also show that for the IoM algorithm 2, $C_{N,M}=L_{N,M}2^N$ with $L_{N,M}$ being a linear function of $N$ and $M$ respectively.  The HOP method is shown to be easy to substitute.

Another major contribution of this paper is a new family of nonlinear CEFs called SVD-CEF. This family of CEFs is of type A and based on the use of components of singular value decomposition (SVD) of a randomly modulated matrix of $\mathbf{x}$. Based on the empirical evidences shown in this paper, the complexity order to attack a SVD-CEF is $C_{N,M}=P_{N,M}2^{\zeta N}$ where $\zeta>1$ is typically substantially larger than one and increases as $N$ increases. Furthermore, we will show that a quantized SVD-CEF also outperforms the IoM algorithm 2 in terms of noise sensitivity.

Table \ref{Table_Comparison} provides a comparison of several CEFs discussed in this work, where ``No'' in the H.I. column is for ``not hard to invert'', ``Yes'' in the H.I. column is for ``empirically hard to invert'', ``Yes'' in the H.S. column is for ``empirically hard to substitute'', ``No'' in the H.S. column is for ``not hard to substitute'', and the column of ``Attack'' is the complexity of attack. The entry marked as ``-'' is no longer important due to a ``No'' in another column.

\begin{table}
  \centering
    \caption{Comparison of CEFs.}\label{Table_Comparison}
  \begin{tabular}{|c|c|c|c|c|c|c|}
     \hline
      & Ref &Type  & H.I. & H.S.& Attack  \\ \hline
     RP & \cite{Teoh2007}&A  & No & -& -  \\
     DRP & \cite{Yang2010}&A   & No & -& -  \\
     URP & Here &A & No & -& -  \\
     HOP & \cite{Grigoriev2012} &A   & - & No& -  \\
     IoM-1 & \cite{Jin2018} &B   & No & -& -  \\
     IoM-2 & \cite{Jin2018}&B    & Yes & Yes& $L_{N,M}2^N$  \\
     SVD-CEF & Here&A   & Yes & Yes& $P_{N,M}2^{\zeta N}$  \\
     \hline
   \end{tabular}
\end{table}

\vspace {-2mm}
\subsection{The Rest of the Paper}
 In section \ref{sec:linearCEF}, we review a linear family of CEFs, including  RP and  DRP.
 We will also discuss the usefulness of unitary random projection (URP), a useful transformation from the $N$-dimensional real space $\mathcal{R}^N$ to the $N$-dimensional sphere of unit radius $\mathcal{S}^N(1)$. In section \ref{sec:NonlinearPrior}, we review a family of nonlinear CEFs, including HOP and  IoM.
  In section \ref{sec:New}, we present a new family of nonlinear CEFs called SVD-CEF, which is a new development from our prior works in \cite{Hua2020a}-\cite{Hua2020b}. In section \ref{sec:hard_to_invert}, we provide empirical details to explain why the SVD-CEF is hard to attack. In section \ref{sec:statistics}, we provide a statistical analyse of the SVD-CEF. In section \ref{sec:comparison}, we show a detailed comparison of the noise sensitivities of a quantized SVD-CEF and the IoM algorithm 2. The conclusion is given in section \ref{sec:conclusion}.

\section{Linear Family of CEFs}\label{sec:linearCEF}

A family of linear CEFs can be expressed as follows:
\begin{equation}\label{}
  \mathbf{y}=\mathbf{R}_S\mathbf{x}
\end{equation}
where $\mathbf{R}_S$ is a $M\times N$ pseudo-random matrix dependent on a secret key $S$. Let the $i$th $M_i\times 1$ subvector of $\mathbf{y}$ be $\mathbf{y}_i$, and the $i$th $M_i\times N$ block matrix of $\mathbf{R}_S$ be $\mathbf{R}_{S,i}$. Then it follows that
\begin{equation}\label{}
  \mathbf{y}_i=\mathbf{R}_{S,i}\mathbf{x}
\end{equation}
where $i=1,\cdots,I$ and $\sum_{i=1}^I M_i=M$.

\subsection{Random Projection}\label{sec:RP}
The linear family of CEFs includes the random projection (RP) method shown in \cite{Teoh2007} and applied in \cite{Pillai2011}. If $S$ is known, so is $\mathbf{R}_{S,i}$ for all $i$. If $\mathbf{y}_i$ for some $i$ is known/exposed and $\mathbf{R}_{S,i}$ is of the full column rank $N$, then $\mathbf{x}$ is given by $\mathbf{R}_{S,i}^+\mathbf{y}_i=(\mathbf{R}_{S,i}^T\mathbf{R}_{S,i})^{-1}\mathbf{R}_{S,i}^T
\mathbf{y}_1$ where $^+$ denotes pseudo-inverse. If $\mathbf{R}_{S,i}$ is not of full column rank, then $\mathbf{x}$ can be computed from a set of outputs like (for example) $\mathbf{y}_1,\cdots,\mathbf{y}_L$ where $L$ is such that the vertical stack of $\mathbf{R}_{S,1},\cdots,\mathbf{R}_{S,L}$, denoted by $\mathbf{R}_{S,1:L}$, is of the full column rank $N$.

If $S$ is unknown, then a method to compute $\mathbf{x}$ includes a discrete search for the $N_S$ bits of $S$ as follows
\begin{equation}\label{}
  \min_S\min_{\mathbf{x}}\|\mathbf{y}_{1:L}-\mathbf{R}_{S,1:L}\mathbf{x}\|=
  \min_S\|\mathbf{y}_{1:L}-\mathbf{R}_{S,1:L}\mathbf{R}_{S,1:L}^+\mathbf{y}_{1:L}\|
\end{equation}
where $\mathbf{y}_{1:L}$ is the vertical stack of $\mathbf{y}_1,\cdots,\mathbf{y}_L$. The total complexity of the above attack algorithm with unknown key $S$ is $P_{N,M} 2^{N_S}$ with $P_{N,M}$ being a linear function of $\sum_{i=1}^LM_i$ and a cubic function of $N$.

So, RP is not hard to attack (subject to a small $N_S$).

\subsection{Dynamic Random Projection}\label{sec:DRP}
The dynamic random projection (DRP) method proposed in \cite{Yang2010} and also discussed in \cite{Patel2015} can be described by
\begin{equation}\label{eq:yiRsixx}
  \mathbf{y}_i=\mathbf{R}_{S,i,\mathbf{x}}\mathbf{x}
\end{equation}
where $\mathbf{R}_{S,i,\mathbf{x}}$ is the $i$th realization of a random matrix that depends on both $S$ and $\mathbf{x}$. Since $\mathbf{R}_{S,i,\mathbf{x}}$ is discrete, $\mathbf{y}_i$ in \eqref{eq:yiRsixx} is a \emph{locally} linear function of $\mathbf{x}$. (There is a nonzero probability that a small perturbation $\mathbf{w}$ in $\mathbf{x}'=\mathbf{x}+\mathbf{w}$ leads to $\mathbf{R}_{S,i,\mathbf{x}'}$ being substantially different from $\mathbf{R}_{S,i,\mathbf{x}}$. This is not a desirable outcome for biometric templates although the probability may be small.) Two methods were proposed in \cite{Yang2010} to construct $\mathbf{R}_{S,i,\mathbf{x}}$, which were called ``Functions I and II'' respectively. For simplicity of notation, we will now suppress $i$ and $S$ in \eqref{eq:yiRsixx} and write it as
\begin{equation}\label{eq:yiRsixx2}
  \mathbf{y}=\mathbf{R}_{\mathbf{x}}\mathbf{x}
\end{equation}

\subsubsection{Assuming ``Function I'' in \cite{Yang2010}}
In this case, the $i$th element of $\mathbf{y}$, denoted by $v_i$, corresponds to the $i$th slot shown in \cite{Yang2010} and can be written as
\begin{equation}\label{}
  v_i=\mathbf{r}_{x,i}^T\mathbf{x}
\end{equation}
where $\mathbf{r}_{x,i}^T$ is the $i$th row of $\mathbf{R}_\mathbf{x}$. But $\mathbf{r}_{x,i}^T$ is one of $L$ key-dependent pseudo-random vectors $\mathbf{r}_{i,1}^T,\cdots,\mathbf{r}_{i,L}^T$ that are independent of $\mathbf{x}$ and known if $S$ is known. So we can also write
\begin{equation}\label{}
  v_i=\mathbf{r}_i^T\mathbf{\bar x}
\end{equation}
where $\mathbf{r}_i^T=[\mathbf{r}_{i,1}^T,\cdots,\mathbf{r}_{i,L}^T]^T$, and $\mathbf{\bar x}\in \mathcal{R}^{LN}$ is a sparse vector consisting of zeros and $\mathbf{x}$. Before $\mathbf{x}$ is known, the position of $\mathbf{x}$ in $\mathbf{\bar x}$ is initially unknown.

If an attacker has stolen $K$ realizations of $v_i$ (denoted by $v_{i,1},\cdots,v_{i,K}$), then it follows that
\begin{equation}\label{}
  \mathbf{v}_i  =\mathbf{R}_i\mathbf{\bar x}
\end{equation}
where $\mathbf{v}_i = [v_{i,1},\cdots,v_{i,K}]^T$, and $\mathbf{R}_i$ is the vertical stack of $K$ key-dependent random realizations of $\mathbf{r}_i^T$. With $K\geq LN$, $\mathbf{R}_i$ is of the full column rank $LN$ with probability one, and in this case the above equation (when given the key $S$) is linearly invertible with a complexity order equal to $\mathcal{O}((LN)^3)$.

An even simpler method of attack is as follows. Since $v_{i,k}=\mathbf{r}_{i,k,l}^T\mathbf{x}$ where $l\in\{1,\cdots,L\}$ and $\mathbf{r}_{i,k,l}$ for all $i$, $k$ and $l$ are known, then we can compute
\begin{align}\label{}
  l^*&=arg\min_{l\in\{1,\cdots,L\}}\min_{\mathbf{x}}\|\mathbf{v}_i-\mathbf{R}_{i,l}\mathbf{x}\|^2\notag\\
  &=
  arg\min_{l\in\{1,\cdots,L\}}\|\mathbf{v}_i-\mathbf{R}_{i,l}\mathbf{R}_{i,l}^+\mathbf{v}_i\|^2
\end{align}
where $\mathbf{R}_{i,l}$ is the vertical stack of $\mathbf{r}_{i,k,l}^T$ for $k=1,\cdots,K$. Provided $K\geq N$, $\mathbf{R}_{i,l}$ has the full column rank with probability one. In this case, the correct solution of $\mathbf{x}$ is given by $\mathbf{R}_{i,l^*}^+\mathbf{v}_i$. This method has a complexity order equal to $\mathcal{O}(LN^3)$.

\subsubsection{Assuming ``Function II'' in \cite{Yang2010}}

To attack ``Function II'' with known $S$, it is equivalent to consider the following signal model:
\begin{equation}\label{}
  v_k = \sum_{n=1}^N r_{k,l_k,n} x_n
\end{equation}
where $v_k$ is available for $k=1,\cdots,K$, $r_{k,l,n}$ for $1\leq k \leq K$, $1\leq l \leq L$ and $1\leq n \leq N$ are random but known\footnote{``random but known'' means ``known'' strictly speaking despite a pseudo-randomness.} numbers (when given $S$), $x_n$ for all $n$ are unknown, and $l_k$ is a $k$-dependent random/unknown choice from $[1,\cdots,L]$.

We can write
\begin{equation}\label{}
  \mathbf{v}=\mathbf{R}\mathbf{x}
\end{equation}
where $\mathbf{v}$ is a stack of all $v_k$, $\mathbf{x}$ is a stack of all $x_n$, and $\mathbf{R}$ is a stack of all $r_{k,l_k,n}$ (i.e., $(\mathbf{R})_{k,n}=r_{k,l_k,n}$). In this case, $\mathbf{R}$ is a random and unknown choice from $L^K$ possible known matrices. An exhaustive search would require the $\mathcal{O}(L^K)$ complexity with $K\geq N+1$.

Now we consider a different approach of attack. Since $r_{k,l,n}$ for all $k,l,n$ are known, we can compute
\begin{equation}\label{}
  c_{n,n'}=\frac{1}{KL}\sum_{k=1}^K\sum_{l=1}^L\sum_{l'=1}^L r_{k,l,n}r_{k,l',n'}
\end{equation}
If $r_{k,l,n}$ are pseudo i.i.d. random (but known) numbers of zero mean and variance one, then for large $K$ (e.g., $K\gg L^2$) we have $c_{n,n'}\approx \delta_{n,n'}$.

Also define
\begin{equation}\label{}
  y_n = \frac{1}{K}\sum_{k=1}^K \sum_{l=1}^L v_k r_{k,l,n} = \sum_{n'=1}^N \hat c_{n,n'} x_{n'}
\end{equation}
where $n=1,\cdots,N$ and
\begin{equation}\label{}
  \hat c_{n,n'} = \frac{1}{K}\sum_{k=1}^K \sum_{l=1}^L r_{k,l,n}r_{k,l_k,n'} .
\end{equation}
If $r_{k,l,n}$ are i.i.d. of zero mean and unit variance, then for large $K$ we have $\hat c_{n,n'}\approx c_{n,n'}\approx \delta_{n,n'}$ and hence
\begin{equation}\label{}
  y_n \approx x_n.
\end{equation}

More generally, if we have $\hat c_{n,n'}\approx c_{n,n'}$ with a large $K$, then
\begin{equation}\label{}
  \mathbf{y}\approx \mathbf{C}\mathbf{x}
\end{equation}
where $(\mathbf{y})_n=y_n$, and $(\mathbf{C})_{n,n'}=c_{n,n'}$. Hence,
\begin{equation}\label{}
  \mathbf{x}\approx \mathbf{C}^{-1}\mathbf{y}.
\end{equation}

With an initial estimate $\mathbf{\hat x}$ of $\mathbf{x}$, we can then do the following to refine the estimate:
\begin{enumerate}
  \item For each of $k=1,\cdots,K$, compute $l_k^*=arg \min_{l\in [1,\cdots,L]} |v_k-\sum_{n=1}^N r_{k,l,n} \hat x_n|$.
  \item Recall $\mathbf{v} = \mathbf{R}\mathbf{x}$. But now use $(\mathbf{R})_{k,n}=r_{k,l_k^*,n}$ for all $k$ and $n$, and replace $\mathbf{\hat x}$ by
      \begin{equation}\label{}
        \mathbf{\hat x}=(\mathbf{R}^T\mathbf{R})^{-1}\mathbf{R}^T\mathbf{v}
      \end{equation}
  \item Go to step 1 until convergence.
\end{enumerate}

Note that all entries in $\mathbf{R}$ are discrete. Once the correct $\mathbf{R}$ is found, the exact $\mathbf{x}$ is obtained. The above algorithm converges to either the exact $\mathbf{x}$ or a wrong $\mathbf{x}$. But with a sufficiently large $K$ with respect to a given pair of $N$ and $L$, our simulation shows that above attack algorithm yields the exact $\mathbf{x}$ with high probabilities. For example, for $N=8$, $L=8$ and $K=23L$, the successful rate is $99\%$. And for $N=16$, $L=48$ and $K=70L$, the successful rate is $98\%$. In the experiment, for each set of $N$, $L$ and $K$, 100 independent realizations of all elements in $\mathbf{x}$ and $\mathbf{R}$ were chosen from i.i.d. Gaussian distribution with zero mean and unit variance. The successful rate was based on the 100 realizations.

In \cite{Yang2010}, an element-wise quantized version of $\mathbf{v}$ was further suggested to improve the hardness to invert. In this case, the vector potentially exposable to an attacker can be written as
\begin{equation}\label{}
  \mathbf{\hat v}=\mathbf{R}\mathbf{x}+\mathbf{w}
\end{equation}
where $\mathbf{w}$ can be modelled as a white noise vector uncorrelated with $\mathbf{R}\mathbf{x}$. The above attack algorithm with $\mathbf{v}$ replaced by $\mathbf{\hat v}$ also applies although a larger $K$ is needed to achieve the same rate of successful attack.

In all of the above cases, the computational complexity for a successful attack is a polynomial function $N$, $L$ and/or $K$ when the secret key $S$ is given.

\subsection{Unitary Random Projection (URP)}\label{sec:URP}
None of the RP and DRP methods is homomorphic.
To have a homomorphic CEF whose input and output have the same distance measure, we can use
\begin{equation}\label{}
  \mathbf{y}_k=\mathbf{R}_k\mathbf{x}
\end{equation}
where $\mathbf{R}_k\in \mathcal{R}^{N\times N}$ for each realization index $k$ is a pseudo-random unitary matrix governed by a secret key $S$. Clearly, if $\mathbf{y}_k'=\mathbf{R}_k\mathbf{x}'$, then $\|\mathbf{y}_k'-\mathbf{y}_k\|=\|\mathbf{x}_k'-\mathbf{x}_k\|$. Clearly, the sensitivity of URP equals one everywhere.

%

If $\mathbf{R}_k$ is just a permutation matrix, then the distribution of the elements of $\mathbf{x}$ is the same as that of $\mathbf{y}_k$ for each $k$.
To hide the distribution of the entries of $\mathbf{x}$ from $\mathbf{y}_k$ for any $k$, we can let $\mathbf{R}_k=\mathbf{P}_{k,2}\mathbf{Q}\mathbf{P}_{k,1}$ where $\mathbf{Q}$ is a fixed unitary matrix (such as the discrete Fourier transform matrix), and $\mathbf{P}_{k,1}$ and $\mathbf{P}_{k,2}$ are pseudo-random permutation matrices governed by the seed $S$. This projection makes the distribution of the elements of $\mathbf{y}_k$ differ from that of $\mathbf{x}$. For large $N$, the distribution of the elements of $\mathbf{y}_k$ approaches the Gaussian distribution for each typical $\mathbf{x}$. Conditioned on a fixed key $S$, if the entries in $\mathbf{x}$ are i.i.d. Gaussian with zero mean and variance $\sigma_x^2$, then the entries in each $\mathbf{y}_i$ are also i.i.d. Gaussian with zero mean and the variance $\sigma_x^2$.

To further scramble the distribution of $\mathbf{y}_k$, we can add one or more layers of pseudo-random permutation and unitary transform, e.g., $\mathbf{R}_k=\mathbf{P}_{k,3}\mathbf{Q}\mathbf{P}_{k,2}\mathbf{Q}\mathbf{P}_{k,1}$.

For unitary $\mathbf{R}_k$, we also have $\|\mathbf{y}_k\|=\|\mathbf{x}\|$, which means that $\|\mathbf{x}\|$ is not protected from $\mathbf{y}_k$. If $\|\mathbf{x}\|$ needs to be protected, we can apply the transformation shown next.

\subsubsection{Transformation from $\mathcal{R}^N$ to $\mathcal{S}^N(1)$}\label{sec:transformation}
 We now introduce a transformation from the $N$-dimensional vector space $\mathcal{R}^N$ to the $N$-dimensional sphere of unit radius $\mathcal{S}^N(1)$.
 Let $\mathbf{x}\in \mathcal{R}^N$. Define
\begin{equation}\label{}
  \mathbf{v}=\left [\begin{array}{c}
                      \frac{1}{\|\mathbf{x}\|\sqrt{1+\|\mathbf{x}\|^2}}\mathbf{x} \\
                      \frac{\|\mathbf{x}\|}{\sqrt{1+\|\mathbf{x}\|^2}}
                    \end{array}
   \right ]
\end{equation}
which clearly satisfies $\mathbf{v}\in\mathcal{S}^N(1)$. Then, we let
\begin{equation}\label{}
  \mathbf{y}_k=\mathbf{R}_k\mathbf{v}
\end{equation}
where $\mathbf{R}_k$ is now a $(n+1)\times (n+1)$ unitary random matrix governed by a secret key $S$.

Let $\mathbf{y}_k'=\mathbf{R}_k\mathbf{v}'$. It follows that $\|\mathbf{y}_k'-\mathbf{y}_k\|=\|\mathbf{v}'-\mathbf{v}\|$. But since $\mathbf{v}$ is now a nonlinear function of $\mathbf{x}$, the relationship between $\|\mathbf{v}'-\mathbf{v}\|$ and $\|\mathbf{x}'-\mathbf{x}\|$ is more complicated, which we discuss below.

Let us consider $\mathbf{x}'=\mathbf{x}+\mathbf{w}$. One can verify that
\begin{align}\label{}
  \|\mathbf{v}'-\mathbf{v}\|&=\left \|\left [ \begin{array}{c}
                                        \frac{\mathbf{x}+
     \mathbf{w}}{\|\mathbf{x}+\mathbf{w}\|\sqrt{1+\|\mathbf{x}+\mathbf{w}\|^2}} \\
      \frac{\|\mathbf{x}+\mathbf{w}\|}{\sqrt{1+\|\mathbf{x}+\mathbf{w}\|^2}}
                                      \end{array}
                                      \right ]
                                      -
                                      \left [ \begin{array}{c}
                                        \frac{\mathbf{x}}{\|\mathbf{x}\|\sqrt{1+\|\mathbf{x}\|^2}} \\
      \frac{\|\mathbf{x}\|}{\sqrt{1+\|\mathbf{x}\|^2}}
                                      \end{array}
                                      \right ]
  \right \|\notag\\
  &= \left \| \left [\begin{array}{c}
                   \frac{\mathbf{a}}{b} \\
                   \frac{c}{d}
                 \end{array}
   \right ]\right \|
\end{align}
where
\begin{align}\label{}
  \mathbf{a}&=(\mathbf{x}+\mathbf{w})\cdot\|\mathbf{x}\|\cdot\sqrt{1+\|\mathbf{x}\|^2}\notag\\
  &-
  \mathbf{x}\cdot\|\mathbf{x}+\mathbf{w}\|\cdot
  \sqrt{1+\|\mathbf{x}+\mathbf{w}\|^2}
\end{align}
\begin{equation}\label{}
  b=\|\mathbf{x}\|\cdot\sqrt{1+\|\mathbf{x}\|^2}\cdot\|\mathbf{x}+\mathbf{w}\|\cdot
  \sqrt{1+\|\mathbf{x}+\mathbf{w}\|^2}
\end{equation}
\begin{equation}\label{}
  c=\|\mathbf{x}+\mathbf{w}\|\cdot\sqrt{1+\|\mathbf{x}\|^2}-\|\mathbf{x}\|\cdot
  \sqrt{1+\|\mathbf{x}+\mathbf{w}\|^2}
\end{equation}
\begin{equation}\label{}
  d=\sqrt{1+\|\mathbf{x}\|^2}\|\cdot\sqrt{1+\|\mathbf{x}+\mathbf{w}\|^2}.
\end{equation}

To derive a simpler relationship between $\|\mathbf{v}'-\mathbf{v}\|$ and $\|\mathbf{x}'-\mathbf{x}\|=\|\mathbf{w}\|$, we will assume $\|\mathbf{w}\|\ll r\doteq \|\mathbf{x}\|$ and apply the first order approximations.
Also we can write
\begin{equation}\label{}
  \mathbf{w} = \eta_x\mathbf{w}_x+\eta_\perp\mathbf{w}_\perp
\end{equation}
where $\mathbf{w}_x$ is a unit-norm vector in the direction of $\mathbf{x} $, and $\mathbf{w}_\perp$ is a unit-norm vector orthogonal to $\mathbf{x}$. Then,
\begin{equation}\label{}
  \|\mathbf{w}\|^2 = \eta_x^2 +\eta_\perp^2
\end{equation}
\begin{equation}\label{}
  \mathbf{x}^T\mathbf{w} = \eta_x\|\mathbf{x}\|=\eta_x r.
\end{equation}

It follows that
\begin{align}\label{}
  \|\mathbf{x}+\mathbf{w}\|&\approx \|\mathbf{x}\|\notag\\
  &+\frac{1}{2\|\mathbf{x}\|}(\|\mathbf{w}\|^2 +2\mathbf{x}^T\mathbf{w})\notag\\
  &=r+\frac{1}{2r}(\eta_x^2 +\eta_\perp^2+2r\eta_x)\notag\\
  &\approx r+\frac{1}{2r}(\eta_\perp^2+2r\eta_x)
\end{align}
\begin{align}\label{}
  \sqrt{1+\|\mathbf{x}+\mathbf{w}\|^2}&\approx \sqrt{1+\|\mathbf{x}\|^2}\notag\\
  &+\frac{1}{2\sqrt{1+\|\mathbf{x}\|^2}}(\|\mathbf{w}\|^2 +2\mathbf{x}^T\mathbf{w})\notag\\
  &\approx\sqrt{1+r^2} +\frac{1}{2\sqrt{1+r^2}}(\eta_\perp^2+2r\eta_x).
\end{align}
Then, one can verify that
\begin{equation}\label{}
  \mathbf{a} \approx \mathbf{w}r\sqrt{1+r^2}-\mathbf{x}\frac{1}{2}\left (\frac{r}{\sqrt{1+r^2}}+\frac{\sqrt{1+r^2}}{r}\right )
  (\eta_\perp^2+2r\eta_x)
\end{equation}
and
\begin{align}\label{}
  \|\mathbf{a}\|^2 &= r^2(1+r^2)(\eta_x^2+\eta_\perp^2)\notag\\
  &+\frac{1}{4}r^2\left (\frac{r}{\sqrt{1+r^2}}
  +\frac{\sqrt{1+r^2}}{r}\right )^2(\eta_\perp^2+2r\eta_x)^2\notag\\
  &-\eta_x r^2\sqrt{1+r^2}\left (\frac{r}{\sqrt{1+r^2}}+\frac{\sqrt{1+r^2}}{r}\right )
  (\eta_\perp^2+2r\eta_x)\notag\\
  &\approx r^2(1+r^2)(\eta_x^2+\eta_\perp^2)\notag\\
  &+r^4\left (\frac{r}{\sqrt{1+r^2}}+\frac{\sqrt{1+r^2}}{r}\right )^2 \eta_x^2\notag\\
  &-2r^3\sqrt{1+r^2}\left (\frac{r}{\sqrt{1+r^2}}+\frac{\sqrt{1+r^2}}{r}\right )
  \eta_x^2\notag\\
  &= r^2(1+r^2) \eta_\perp^2 +\frac{r^6}{1+r^2}\eta_x^2
\end{align}
where the approximations hold because of $\eta_x\ll r$ and $\eta_\perp\ll r$.
Similarly, we have
\begin{equation}\label{}
  b^2\approx r^4(1+r^2)^2
\end{equation}
\begin{equation}\label{}
  c^2\approx \left (\frac{1}{2r\sqrt{1+r^2}}(\eta_\perp^2+2r\eta_x)\right )^2\approx
  \frac{1}{(1+r^2)}\eta_x^2
\end{equation}
\begin{equation}\label{}
  d^2 \approx (1+r^2)^2.
\end{equation}
Hence
\begin{equation}\label{}
  \|\mathbf{v}'-\mathbf{v}\|^2 =\frac{\|\mathbf{a}\|^2}{b^2}+\frac{c^2}{d^2}
  \approx\frac{1}{r^2(1+r^2)} \eta_\perp^2 +\frac{r^2+1}{(1+r^2)^3}\eta_x^2.
\end{equation}

It is somewhat expected that the larger is $r$, the less are the sensitivities of $\|\mathbf{v}'-\mathbf{v}\|^2$ to $\eta_\perp$ and $\eta_x$.
But the sensitivities of $\|\mathbf{v}'-\mathbf{v}\|^2$ to $\eta_\perp$ and $\eta_x$ are different in general, which also vary differently as $r$ varies. If $r\ll 1$, then
\begin{equation}\label{}
  \|\mathbf{v}'-\mathbf{v}\|^2 \approx \frac{1}{r^2} \eta_\perp^2 +\eta_x^2
\end{equation}
which shows a higher sensitivity of $\|\mathbf{v}'-\mathbf{v}\|^2$  to $\eta_\perp$ than to $\eta_x$.
If $r\gg 1$, then
\begin{equation}\label{eq:vvw}
  \|\mathbf{v}'-\mathbf{v}\|^2
  \approx\frac{1}{r^4} \eta_\perp^2 +\frac{1}{r^4}\eta_x^2=\frac{1}{r^4}\|\mathbf{w}\|^2
\end{equation}
which shows equal sensitivities  of $\|\mathbf{v}'-\mathbf{v}\|^2$ to $\eta_\perp$ and $\eta_x$ respectively.

The above results show how $\|\mathbf{v}'-\mathbf{v}\|^2$ changes with $\mathbf{w}=\eta_\perp \mathbf{w}_\perp +\eta_x \mathbf{w}_x$ subject to $\|\mathbf{w}\|\ll \|\mathbf{x}\|=r$ or equivalently $\sqrt{\eta_\perp^2+\eta_x^2}\ll r$.

For larger $\|\mathbf{w}\|$, the relationship between $\|\mathbf{v}'-\mathbf{v}\|^2$ and $\|\mathbf{w}\|$ is not as simple. But one can verify that if $\|\mathbf{w}\|\gg r \gg 1$, then $\|\mathbf{v}'-\mathbf{v}\|\approx 1/r$.

\section{Nonlinear Family of CEFs}\label{sec:NonlinearPrior}

If the secret key $S$ available is not large enough, then we must consider CEFs that cannot be proven to be easy to attack even if $S$ is known. Such a CEF has to be nonlinear.

\subsection{Higher-Order Polynomials}
A family of higher-order polynomials (HOP) was suggested in \cite{Grigoriev2012} as a hard-to-invert continuous function. But we show next that HOP is not hard to substitute. Let $\mathbf{y}=[y_1,\cdots,y_M]^T$ and $\mathbf{x}=[x_1,\cdots,x_N]^T$ where $y_m$ is a HOP of $x_1,\cdots,x_N$ with pseudo-random coefficients. Namely,  $y_m=f_m(x_1,\cdots,x_N)=\sum_{j=0}^J c_{m,j} x_1^{p_{1,j}}\cdots x_N^{p_{N,j}}$ where the coefficients $c_{m,j}$ are pseudo-random numbers governed by $S$. When $S$ is known, all the polynomials are known and yet $\mathbf{x}$ is still generally hard to obtain from $\mathbf{y}$ for any $M$ due to the nonlinearity. But we can write $y_m=g_m(\mathbf{v}(x_1,\cdots,x_N))$, where $g_m$ is a scalar linear function conditioned on $S$, and $\mathbf{v}(x_1,\cdots,x_N)$ is a $J\times 1$ vector nonlinear function unconditioned on $S$. This means that the HOP is not a hard-to-substitute function.

\subsection{Index-of-Max Hashing}

More recently a method called index-of-max (IoM) hashing was proposed in \cite{Jin2018} and applied in \cite{Kirchgasser2020}. There are algorithms 1 and 2 based on IoM, which will be referred to as IoM-1 and IoM-2.

In IoM-1, the feature vector $\mathbf{x}\in \mathcal{R}^N$ is multiplied (from the left) by a sequence of $L\times N$ pseudo-random matrices $\mathbf{R}_1,\cdots,\mathbf{R}_{K_1}$ to produce $\mathbf{v}_1,\cdots,\mathbf{v}_{K_1}$ respectively. The index of the largest element in each $\mathbf{v}_k$ is used as an output $y_k$. With $\mathbf{y}=[y_1,\cdots,y_{K_1}]^T$, we see that $\mathbf{y}$ is a nonlinear (``piece-wise'' constant and ``piece-wise'' continuous) continuous function of $\mathbf{x}$.

In IoM-2, $\mathbf{R}_1,\cdots,\mathbf{R}_{K_1}$ used in IoM-1 are replaced by $N\times N$ pseudo-random permutation matrices $\mathbf{P}_1,\cdots,\mathbf{P}_{K_1}$ to produce $\mathbf{v}_1,\cdots,\mathbf{v}_{K_1}$, and then a sequence of vectors $\mathbf{w}_1,\cdots,\mathbf{w}_{K_2}$ are produced in such a way that each $\mathbf{w}_k$ is the element-wise products of an exclusive set of $p$ vectors from $\mathbf{v}_1,\cdots,\mathbf{v}_{K_1}$. The index of the largest element in each $\mathbf{w}_k$ is used as an output $y_k$. With $\mathbf{y}=[y_1,\cdots,y_{K_2}]^T$, we see that $\mathbf{y}$ is another nonlinear continuous function of $\mathbf{x}$.

Next we show that IoM-1 is not hard to invert if the secret key $S$ or equivalently the random matrices $\mathbf{R}_1,\cdots,\mathbf{R}_{K_1}$ are known. We also show that IoM-2 is not hard to invert up to the sign of each element in $\mathbf{x}$ if the secret key $S$ or equivalently the random permutations $\mathbf{R}_1,\cdots,\mathbf{R}_{K_1}$ are known.

\subsubsection{Attack of IoM-1}
Assume that each $\mathbf{R}_k$ has $L$ rows and the secret key $S$ is known. Then knowing $y_k$ for $k=1,\cdots,K_1$ means knowing $\mathbf{r}_{k,a,l}$ and $\mathbf{r}_{k,b,l}$ satisfying
\begin{equation}\label{}
  \mathbf{r}_{k,a,l}^T\mathbf{x}>\mathbf{r}_{k,b,l}^T\mathbf{x}
\end{equation}
with $l=1,\cdots,L-1$ and $k=1,\cdots,K_1$. Here $\mathbf{r}_{k,a,l}^T$ and $\mathbf{r}_{k,b,l}^T$ for all $l$ are rows of $\mathbf{R}_k$. The above is equivalent to
$
  \mathbf{d}_{k,l}^T\mathbf{x}>0
$
with $\mathbf{d}_{k,l}=\mathbf{r}_{k,a,l}-\mathbf{r}_{k,b,l}$, or more simply
\begin{equation}\label{eq:dkx}
  \mathbf{d}_k^T\mathbf{x}>0
\end{equation}
where $\mathbf{d}_k$ is known for $k=1,\cdots,K$ with $K=K_1(L-1)$. Note that any scalar change to $\mathbf{x}$ does not affect the output $\mathbf{y}$. Also note that even though IoM-1 defines a nonlinear function from $\mathbf{x}$ to $\mathbf{y}$, the conditions in \eqref{eq:dkx} useful for attack are linear with respect to $\mathbf{x}$.

To attack IoM-1, we can simply compute $\mathbf{\hat x}$ satisfying $\mathbf{d}_k^T\mathbf{\hat x}>0$ for all $k$. One such algorithm of attack is as follows:
\begin{enumerate}
  \item Initialization/averaging: Let  $\mathbf{\hat x}=\mathbf{\bar d}\doteq\frac{1}{K}\sum_{k=1}^K\mathbf{d}_k$.
  \item Refinement: Until $\mathbf{d}_k^T\mathbf{\hat x}>0$ for all $k$, choose $k^*=arg\min_k \mathbf{d}_k^T\mathbf{\hat x}$, and compute
      \begin{equation}\label{}
        \mathbf{\hat x}\leftarrow\mathbf{\hat x} -\eta (\mathbf{d}_{k^*}^T\mathbf{\hat x})\mathbf{d}_{k^*}
      \end{equation}
      where $\eta$ is a step size.
\end{enumerate}
Our simulation (using $\eta = \frac{1}{\|\mathbf{d}_{k^*}\|^2}$) shows that using the initialization alone can yield a good estimate of $\mathbf{x}$ as $K$ increases. More specifically, the normalized projection $\frac{\mathbf{\bar d}^T\mathbf{x}}{\|\mathbf{\bar d}\|\cdot\|\mathbf{x}\|}$ converges to one as $K$ increases. Our simulation also shows that the second step in the above algorithm improves the convergence slightly. Examples of the attack results are shown in Tables \ref{Table1} and \ref{Table2} where $L=N$. We see that IoM-1 (with its key $S$ exposed) can be inverted with a complexity order no larger than a linear function of $N$ and $K_1$ respectively.

\begin{table}
  \centering
  \caption{Normalized projection of $\mathbf{x}$ onto its estimate using only averaging for attack of IoM-1}\label{Table1}
  \begin{tabular}{|c|c|c|c|c|}
    \hline
     & $K_1=8$ & 16 & 32 & 64 \\
    \hline
    $N=8$ & 0.8546 & 0.9171 & 0.9562 & 0.9772 \\
    16 & 0.8022 & 0.8842 & 0.9365 &0.9666 \\
    32 & 0.7328 & 0.8351 & 0.906 & 0.9494 \\
    \hline
  \end{tabular}
\end{table}

\begin{table}
  \centering
  \caption{Normalized projection of $\mathbf{x}$ onto its estimate after convergence of refinement for attack of IoM-1}\label{Table2}
  \begin{tabular}{|c|c|c|c|c|}
    \hline
     & $K_1=8$ & 16 & 32 & 64 \\
    \hline
    $N=8$ & 0.8807 & 0.9467 & 0.9804 & 0.9937 \\
    16 & 0.8174 & 0.908 & 0.9612 &0.9861 \\
    32 & 0.739 & 0.8497 & 0.9268 & 0.9699 \\
    \hline
  \end{tabular}
\end{table}

\subsubsection{Attack of IoM-2}
To attack IoM-2, we need to know the sign of each element of $\mathbf{x}$, which is assumed below. Given the output of IoM-2 and all the permutation matrices $\mathbf{P}_1,\cdots,\mathbf{P}_{K_1}$, we know which of the elements in each $\mathbf{w}_k$ is the largest and which of these elements are negative. If the largest element in $\mathbf{w}_k$ is positive, we will ignore all the negative elements in $\mathbf{w}_k$. If the largest element in $\mathbf{w}_k$ is negative, we know which of the elements in $\mathbf{w}_k$ has the smallest absolute value.

Let $|\mathbf{w}_k|$ be the vector consisting of the corresponding absolute values of the elements in $\mathbf{w}_k$. Also let $\log|\mathbf{w}_k|$ be the vector of element-wise logarithm of $|\mathbf{w}_k|$. It follows that
\begin{equation}\label{}
  \log|\mathbf{w}_k|=\mathbf{T}_k\log|\mathbf{x}|
\end{equation}
where $\mathbf{T}_k$ is the sum of the permutation matrices used for $\mathbf{w}_k$. The knowledge of an output $y_k$ of IoM-2 implies the knowledge of $\mathbf{t}_{k,a,l}^T$ and $\mathbf{t}_{k,b,l}^T$  (i.e., row vectors of $\mathbf{T}_k$) such that either
\begin{equation}\label{}
  \mathbf{t}_{k,a,l}^T\log|\mathbf{x}|>\mathbf{t}_{k,b,l}\log|\mathbf{x}|
\end{equation}
with $l=1,\cdots,L_k-1$ if $\mathbf{w}_k$ has $L_k\geq 2$ positive elements, or
\begin{equation}\label{}
  \mathbf{t}_{k,a,l}^T\log|\mathbf{x}|<\mathbf{t}_{k,b,l}\log|\mathbf{x}|
\end{equation}
with $l=1,\cdots,N-1$ if $\mathbf{w}_k$ has no positive element.

 If $\mathbf{w}_k$ has only one positive element, the corresponding $y_k$ is ignored as it yields no useful constraint on $\log|\mathbf{x}|$. We assume that no element in $\mathbf{x}$ is zero.

Equivalently, the knowledge of $y_k$ implies
$
  \mathbf{c}_{k,l}^T\log|\mathbf{x}|>0
$
where $\mathbf{c}_{k,l}=\mathbf{t}_{k,a,l}-\mathbf{t}_{k,b,l}$ for $l=1,\cdots,L_k-1$ if $\mathbf{w}_k$ has $L_k\geq 2$ positive elements, or $\mathbf{c}_{k,l}=-\mathbf{t}_{k,a,l}+\mathbf{t}_{k,b,l}$ for $l=1,\cdots,N-1$ if $\mathbf{w}_k$ has no positive element. A simpler form of the constraints on $\log|\mathbf{x}|$ is
\begin{equation}\label{eq:ckx}
  \mathbf{c}_k^T\log|\mathbf{x}|>0
\end{equation}
where $\mathbf{c}_k$ is known for $k=1,\cdots,K$ with $K=\sum_{k=1}^{K_2}(\bar L_k-1)$. Here $\bar L_k = L_k$ if $\mathbf{w}_k$ has a positive element, and $\bar L_k = N$ if $\mathbf{w}_k$ has no positive element.

The algorithm to find $\log|\mathbf{x}|$ satisfying \eqref{eq:ckx} for all $k$ is similar to that for \eqref{eq:dkx}, which consists of ``initialization/averaging'' and ``refinement''. Knowing $\log|\mathbf{x}|$, we also know $|\mathbf{x}|$. Examples of the attack results are shown in Tables \ref{Table3} and \ref{Table4} where $p=N$ and all entries of $\mathbf{x}$ are assumed to be positive.

The above analysis shows that IoM-2 effectively extracts out a binary (sign) secret from each element of $\mathbf{x}$ and utilizes that secret to construct its output. Other than that secret, IoM-2 is not a hard-to-invert function. In other words, IoM-2 can be inverted with a complexity order no larger than $L_{N,K_2}2^N$ where $L_{N,K_2}$ is a linear function of $N$ and $K_2$, respectively, and $2^N$ is to due to an exhaustive search of the sign of each element in $\mathbf{x}$. Note that if an additional key $S_x$ of $N$ bits is first extracted with 100\% reliability from the signs of the elements in $\mathbf{x}$, then a linear CEF could be used while maintaining an attack complexity order equal to $\mathcal{O}(N^3 2^N)$.

\begin{table}
  \centering
  \caption{Normalized projection of $\mathbf{|x|}$ onto its estimate using only averaging for attack of IoM-2}\label{Table3}
  \begin{tabular}{|c|c|c|c|c|}
    \hline
     & $K_2=8$ & 16 & 32 & 64 \\
    \hline
    $N=8$ & 0.9244 & 0.954 & 0.9698 & 0.9783 \\
    16 & 0.9068 & 0.9418 & 0.9603 &0.9694 \\
    32 & 0.8844 & 0.9206 & 0.9379 & 0.9466 \\
    \hline
  \end{tabular}
\end{table}

\begin{table}
  \centering
  \caption{Normalized projection of $\mathbf{|x|}$ onto its estimate after convergence of refinement for attack of IoM-2}\label{Table4}
  \begin{tabular}{|c|c|c|c|c|}
    \hline
     & $K_2=8$ & 16 & 32 & 64 \\
    \hline
    $N=8$ & 0.9432 & 0.9711 & 0.9802 & 0.9816 \\
    16 & 0.9182 & 0.9525 & 0.9649 &0.9653 \\
    32 & 0.8887 & 0.9258 & 0.9403 & 0.9432 \\
    \hline
  \end{tabular}
\end{table}

\section{A New Family of Nonlinear CEFs}\label{sec:New}

The previous discussions show that RP, DRP and IoM-1 are not hard to invert, and IoM-2 can be inverted with a complexity order no larger than $L_{N,K_2}2^N$.
We show next a new family of nonlinear CEFs, for which the best known method to attack suffers a complexity order no less than  $\mathcal{O}(2^{\zeta N})$ with $\zeta$ substantially larger than one.

The new family of nonlinear CEFs is broadly defined as follows. Step 1: let $\mathbf{M}_{k,x}$ be a matrix (for index $k$) consisting of elements that result from a random modulation of the input vector $\mathbf{x}\in\mathcal{R}^N$. Step 2: Each element of the output vector $\mathbf{y}\in\mathcal{R}^M$ is constructed from a component of the singular value decomposition (SVD) of $\mathbf{M}_{k,x}$ for some $k$. Each of the two steps can have many possibilities.  We will next focus on one specific CEF in this family (as this CEF seems the best among many choices we have considered).

 For each pair of $k$ and $l$, let $\mathbf{Q}_{k,l}$ be a (secret key dependent) random  $N\times N$ unitary (real) matrix.
Define
\begin{equation}\label{eq:Mkx}
  \mathbf{M}_{k,x}=[\mathbf{Q}_{k,1}\mathbf{x},\cdots,\mathbf{Q}_{k,N}\mathbf{x}]
\end{equation}
where each column of $\mathbf{M}_{k,x}$ is a random rotation of  $\mathbf{x}$.
Let $\mathbf{u}_{k,x,1}$ be the principal left singular vector of $\mathbf{M}_{k,x}$, i.e.,
\begin{equation}\label{eq:typeA}
  \mathbf{u}_{k,x,1}=arg\max_{\mathbf{u},\|\mathbf{u}\|=1}
  \mathbf{u}^T\mathbf{M}_{k,x}\mathbf{M}_{k,x}^T\mathbf{u}
\end{equation}
Then for each $k$, choose $N_y<N$ elements in $\mathbf{u}_{k,x,1}$ to be $N_y$ elements in  $\mathbf{y}$. For convenience, we will refer to the above function (from $\mathbf{x}$ to $\mathbf{y}$) as SVD-CEF. Note that there are efficient ways to perform the forward computation needed for \eqref{eq:typeA} given $\mathbf{M}_{k,x}\mathbf{M}_{k,x}^T$. One of them is the power method \cite{Golub}, which has the complexity equal to $\mathcal{O}(N^2)$. But the construction of $\mathbf{M}_{k,x}\mathbf{M}_{k,x}^T$ requires $\mathcal{O}(N^3)$ complexity.

We can see that for each random realization of $\mathbf{Q}_{k,l}$ for all $k$ and $l$ and a random realization $\mathbf{x}_0$ of $\mathbf{x}$, with probability one there is a neighborhood around $\mathbf{x}_0$ within which $\mathbf{y}$ is a continuous function of $\mathbf{x}$.
It is also clear that for any fixed $\mathbf{x}$ the elements in $\mathbf{y}$ appear random to anyone who does not have access to the secret key used to produce the pseudo-random $\mathbf{Q}_{k,l}$.

In the next two sections, we will provide detailed analyses  of the SVD-CEF.

\section{Attack of the SVD-CEF}\label{sec:hard_to_invert}

We now consider how to compute $\mathbf{x}\in \mathcal{R}^N$ from a given $\mathbf{y}\in \mathcal{R}^M$ with $M\geq N$ for the SVD-CEF based on \eqref{eq:Mkx} and \eqref{eq:typeA} assuming that $\mathbf{Q}_{k,l}$ for all $k$ and $l$ are also given.

One method (a universal method) is via exhaustive search in the space of $\mathbf{x}$ until a desired $\mathbf{x}$ is found (which produces the known $\mathbf{y}$ via the forward function).   This method has a complexity order (with respect to $N$) no less than $\mathcal{O}(2^{N_BN})$ with $N_B$ being the number of bits needed to represent each element in $\mathbf{x}$. The value of $N_B$ depends on noise level in $\mathbf{x}$. It is not uncommon in practice that $N_B$ ranges from 3 to 8 or even larger.

The only other good method to invert the SVD-CEF seems the Newton's method, which is considered next. To prepare for the application of the Newton's method, we need to formulate a set of equations which must be satisfied by all unknown variables.

\subsection{Preparation}
 We now assume that for each of $k=1,\cdots,K$, $N_y$ elements of $\mathbf{u}_{k,x,1}$ are used to construct $\mathbf{y}\in \mathcal{R}^M$ with $M=KN_y$.  To find $\mathbf{x}$ from known $\mathbf{y}$ and known $\mathbf{Q}_{k,l}$ for all $k$ and $l$, we can solve the following eigenvalue-decomposition (EVD) equations:
\begin{equation}\label{eq:MMyk}
  \mathbf{M}_{k,x}\mathbf{M}_{k,x}^T\mathbf{u}_{k,x,1}=\sigma_{k,x,1}^2\mathbf{u}_{k,x,1}
\end{equation}
with $k=1,\cdots,K$.  Here $\sigma_{k,x,1}^2$ is the principal eigenvalue of $\mathbf{M}_{k,x}\mathbf{M}_{k,x}^T$. But this is not a conventional EVD problem because the vector $\mathbf{x}$ inside $\mathbf{M}_{k,x}$ is unknown along with $\sigma_{k,x,1}^2$ and $N-N_y$ elements in $\mathbf{u}_{k,x,1}$ for each $k$. We will refer to \eqref{eq:MMyk} as the EVD equilibrium conditions for $\mathbf{x}$.

If the unknown $\mathbf{x}$ is multiplied by $\alpha$, so should be the corresponding unknowns $\sigma_{k,x,1}$ for all $k$ but $\mathbf{u}_{k,x,1}$ for any $k$ is not affected.
So, we will only need to consider the solution satisfying $\|\mathbf{x}\|^2=1$. Note that if the norm of the original feature vector contains secret, we can first use the transformation shown in section \ref{sec:transformation}.

The number of unknowns in the system of nonlinear equations \eqref{eq:MMyk} is $N_{unk,EVD,1}=N+(N-N_y)K+K$, which consists of all $N$ elements of $\mathbf{x}$, $N-N_y$ elements of $\mathbf{u}_{k,x,1}$ for each $k$ and $\sigma_{k,x,1}^2$ for all $k$. The number of the nonlinear equations is $N_{equ,EVD,1}=NK+K+1$, which consists of \eqref{eq:MMyk} for all $k$, $\|\mathbf{u}_{k,x,1}\|=1$ for all $k$ and $\|\mathbf{x}\|^2=1$. Then, the necessary condition for a finite set of solutions is $N_{equ,EVD,1}\geq N_{unk,EVD,1}$, or equivalently $N_yK\geq N-1$.

If $N_y<N$, there are $N-N_y$ unknowns in $\mathbf{u}_{k,x,1}$ for each $k$ and hence the left side of \eqref{eq:MMyk} is a third-order function of unknowns. To reduce the nonlinearity, we can expand the space of unknowns as follows. Since $\mathbf{M}_{k,x}\mathbf{M}_{k,x}^T=\sum_{l=1}^N\mathbf{Q}_{k,l}\mathbf{X}\mathbf{Q}_{k,l}^T$ with $\mathbf{X}=\mathbf{x}\mathbf{x}^T$, we can treat $\mathbf{X}$ as a $N\times N$ symmetric unknown matrix (without the rank-1 constraint), and rewrite \eqref{eq:MMyk} as
\begin{equation}\label{eq:QXKyk}
  (\sum_{l=1}^N\mathbf{Q}_{k,l}\mathbf{X}\mathbf{Q}_{k,l}^T)\mathbf{u}_{k,x,1} = \sigma_{k,x,1}^2\mathbf{u}_{k,x,1}
\end{equation}
with $Tr(\mathbf{X})=1$, $\|\mathbf{u}_{k,x,1}\|=1$ and $k=1,\cdots,K$.
In this case, both sides of \eqref{eq:QXKyk} are of the 2nd order of all unknowns. But the number of unknowns is now $N_{unk,EVD,2}=\frac{1}{2}N(N+1)+(N-N_y)K+K>N_{unk,EVD,1}$ while the number of equations is not changed, i.e., $N_{equ,EVD,2}=N_{equ,EVD,1}=NK+K+1$. In this case, the necessary condition for a finite set of solution for $\mathbf{X}$ is $N_{equ,EVD,2}\geq N_{unk,EVD,2}$, or equivalently $N_yK\geq \frac{1}{2}N(N+1)-1$.

Note that $\mathbf{X}$ seems the only useful substitute for $\mathbf{x}$. But this substitute still seems hard to compute from $\mathbf{y}$ as shown later.

Alternatively, we know that $\mathbf{x}$ satisfies the following SVD equations:
\begin{equation}\label{eq:MUV}
  \mathbf{M}_{k,x}\mathbf{V}_{k,x}=\mathbf{U}_{k,x}\boldsymbol{\Sigma}_{k,x}
\end{equation}
with $\mathbf{U}_{k,x}^T\mathbf{U}_{k,x}=\mathbf{I}_N$ and $\mathbf{V}_{k,x}^T\mathbf{V}_{k,x}=\mathbf{I}_N$.
Here $\mathbf{U}_{k,x}$ is the matrix of all left singular vectors, $\mathbf{V}_{k,x}$ is the matrix of all right singular vectors, and $\boldsymbol{\Sigma}_{k,x}$ is the diagonal matrix of all singular values.
The above equations are referred to as the SVD equilibrium  conditions on $\mathbf{x}$.

With $N_y$ elements of the first column of $\mathbf{U}_{k,x}$ for each $k$ to be known, the unknowns are the vector $\mathbf{x}$, $N^2-N_y$ elements in $\mathbf{U}_{k,x}$ for each $k$, all $N^2$ elements in $\mathbf{V}_{k,x}$ for each $k$, and all diagonal elements in $\boldsymbol{\Sigma}_{k,x}$ for each $k$.
Then, the number of unknowns is now $N_{unk,SVD}=N+(N^2-N_y)K+N^2K+NK$, and the number of equations is $N_{equ,SVD}=N^2K+N(N+1)K+1$.  In this case, $N_{equ,SVD}\geq N_{unk,SVD}$ iff $N_yK\geq N-1$. This is the same condition as that for EVD equilibrium.  But the SVD equilibrium equations in \eqref{eq:MUV} are all of the second order.

Note that for the EVD equilibrium, there is no coupling between different eigen-components. But for the SVD equilibrium, there are couplings among all singular-components. Hence the latter involves a much larger number of unknowns than the former. Specifically, $N_{unk,SVD}>N_{unk,EVD,2}>N_{unk,EVD,1}$.

Every set of equations that $\mathbf{x}$ must fully satisfy (given $\mathbf{y}$) is a set of nonlinear equations, regardless of how the parameterization is chosen. This seems the fundamental reason why the SVD-CEF is hard to invert. SVD is a three-factor decomposition of a real-valued matrix, for which there are efficient ways for forward computations but no easy way for backward computation. If a two-factor decomposition of a real-valued matrix (such as QR decomposition) is used, the hard-to-invert property does not seem achievable.

In Appendix \ref{sec:attack}, the details of an attack algorithm based on Newton's method are given.

\subsection{Performance of Attack Algorithm}\label{sec:performance_of_attack}
 Since the conditions useful for attack of the SVD-CEF are always nonlinear, any attack algorithm with a random initialization $\mathbf{x}'$ can converge to the true vector $\mathbf{x}$ (or its equivalent which produces the same $\mathbf{y}$) only if $\mathbf{x}'$ is close enough to $\mathbf{x}$. To translate the local convergence into a computational complexity needed to successfully obtain $\mathbf{x}$ from $\mathbf{y}$, we now consider the following.

Let $\mathbf{x}$ be an $N$-dimensional unit-norm vector of interest. Any unit-norm initialization of $\mathbf{x}$ can be written as
\begin{equation}\label{eq:xxw}
  \mathbf{x}'=\pm \sqrt{1-r^2}\mathbf{x} + r\mathbf{w}
\end{equation}
where $0<r\leq 1$ and $\mathbf{w}$ is a unit-norm vector orthogonal to $\mathbf{x}$.
For any $\mathbf{x}$, $r\mathbf{w}$ is a vector (or ``point'') on the sphere of dimension $N-2$ and radius $r$, denoted by $\mathcal{S}^{N-2}(r)$.
The total area of $\mathcal{S}^{N-2}(r)$ is known to be $|\mathcal{S}^{N-2}(r)|= \frac{2\pi^{\frac{N-1}{2}}}{\Gamma(\frac{N-1}{2})}r^{N-2}$.
Then the probability for a uniformly random $\mathbf{x}'$ from $\mathcal{S}^{N-1}(1)$ to fall onto $\mathcal{S}^{N-2}(r_0)$ orthogonal to $\sqrt{1-r_0^2}\mathbf{x}$ with $r\leq r_0\leq  r+dr$ is $2\frac{|\mathcal{S}^{N-2}(r)|}{|\mathcal{S}^{N-1}(1)|}dr$ where the factor 2 accounts for $\pm$ in \eqref{eq:xxw}.

Therefore, the probability of convergence from $\mathbf{x}'$  to $\mathbf{x}$  is
\begin{align}\label{}
  P_{conv}&=\mathcal{E}_x\left \{ \int_0^1 2 P_{x,r}\frac{|\mathcal{S}^{N-2}(r)|}{|\mathcal{S}^{N-1}(1)|}dr\right \}\notag\\
  &=\frac{2\Gamma \left (\frac{N}{2}\right )}{\sqrt{\pi}\Gamma \left (\frac{N-1}{2}\right )}\int_0^1P_r r^{N-2}dr
\end{align}
where $\mathcal{E}_x$ is the expectation over $\mathbf{x}$, $P_{x,r}$ is the probability of convergence from $\mathbf{x}'$ to $\mathbf{x}$ when $\mathbf{x}'$ is chosen randomly from $\mathcal{S}^{N-2}(r)$ orthogonal to a given $\sqrt{1-r^2}\mathbf{x}$, and $\mathcal{E}_x\{P_{x,r}\}=P_r$.

We see that $P_r$ is the probability that the algorithm converges from $\mathbf{x}'$  to $\mathbf{x}$ (including its equivalent) subject to a fixed $r$, uniformly random unit-norm $\mathbf{x}$, and uniformly random unit-norm $\mathbf{w}$ satisfying $\mathbf{w}^T\mathbf{x}=0$. And $P_r$ can be estimated via simulation.

If $P_r=0$ for $r\geq r_{max}$ (with $r_{max}<1$), then
\begin{align}\label{}
  P_{conv}&=\frac{2\Gamma \left (\frac{N}{2}\right )}{\sqrt{\pi}\Gamma \left (\frac{N-1}{2}\right )}\int_0^{r_{max}}P_r r^{N-2}dr\notag\\
  &<\frac{2\Gamma \left (\frac{N}{2}\right )}{(N-1)\sqrt{\pi}\Gamma \left (\frac{N-1}{2}\right )}r_{max}^{N-1}\notag\\
  &<r_{max}^{N-1}
\end{align}
which converges to zero exponentially as $N$ increases. In other words, for such an algorithm to find $\mathbf{x}$ or its equivalent from random initializations has a complexity order equal to $\mathcal{O}(\frac{1}{P_{conv}})>\mathcal{O}((\frac{1}{r_{max}})^{N-1})$ which increases exponentially as $N$ increases.

In our simulation, we have found that $r_{max}$ decreases rapidly as $N$ increases. Let $P_{r,N}$ be $P_r$ as function of $N$. Also let $P_{r,N}^*$ be the probability of convergence to $\mathbf{\hat x}$ which via the SVD-CEF not only yields the correct $y_k$ for $k=1,\cdots,K$ but also the correct $y_k$ for $k>K$ (up to maximum absolute element-wise error no larger than 0.02). Here $K$ is the number of output elements used to compute the input vector $\mathbf{x}$. In the simulation, we chose $N_y=1$ and $N_{equ,EVD,2}=N_{unk,EVD,2}+1$, which is equivalent to $K=\frac{1}{2}N(N+1)$. Shown in Table \ref{Table_Pr} are the percentage values of $P_{r,N}$ versus $r$ and $N$, which are based on 100 random choices of $\mathbf{x}$. For each choice of $\mathbf{x}$ and each value of $r$, we used one random initialization of $\mathbf{x}'$. (For $N=8$ and the values of $r$ in this table, it took two days on a PC with CPU 3.4 GHz Dual Core to complete the 100 runs.)

\begin{table}
  \centering
  \caption{$P_{r,N}$ and $P_{r,N}^*$ in $\%$ versus $r$ and $N$}\label{Table_Pr}
\begin{tabular}{ |c|c|c|c|c|c|c|c|c| }
\hline
 $r$ & $0.001$ & $0.01$ & $0.1$ &  $0.3$ &  $0.5$ &  $0.7$& $0.9$ & $1$\\
\hline
$P_{r,4}$ & $46$ & $24$ & $6$ &  $0$ &  $1$ &  $1$  & $1$& $0$ \\
\hline
$P_{r,4}^*$  & $45$ & $17$ & $4$ &  $0$ &  $1$ & $0$  & $1$& $0$\\
\hline
$P_{r,8}$  & $29$ & $7$ & $1$ &  $0$ &  $0$ & $0$  & $0$ & $0$\\
\hline
$P_{r,8}^*$  & $25$ & $5$ & $0$ &  $0$ &  $0$ &  $0$ & $0$ & $0$\\
\hline
\end{tabular}
\end{table}


\section{Statistics of the SVD-CEF}\label{sec:statistics}

In this section, we show a statistical study of the SVD-CEF, which shows how sensitive the SVD-CEF is in terms of its input perturbation versus its output perturbation. We will also show a weak correlation between its input and output, and a weak correlation between its outputs. These weak correlations are certainly desirable for a CEF.

The statistics of the output $\mathbf{y}$ of the SVD-CEF is directly governed by the statistics of the principal eigenvector $\mathbf{u}_k\doteq\mathbf{u}_{k,x,1}$ of the matrix $\mathbf{M}_{k,x}\mathbf{M}_{k,x}^T$. So, we can next focus on the statistics of $\mathbf{u}_k$ versus $\mathbf{x}$.

\subsection{Input-Output Distance Relationships}
Unlike the random unitary projections, here the relationship between $\|\Delta \mathbf{x}\|$ and $\Delta \mathbf{u}_k$ is much more complicated.

\subsubsection{Local Sensitivities}\label{sec:local}
For local sensitivities, we consider the relationship between the differentials $\partial \mathbf{u}_{k,x,1}$ versus $\partial \mathbf{x}$.

Since $\mathbf{u}_{k,x,1}$ is the principal eigenvector of $\mathbf{M}_{k,x}\mathbf{M}_{k,x}^T=\sum_{l=1}^N\mathbf{Q}_{k,l}\mathbf{x}\mathbf{x}^T\mathbf{Q}_{k,l}^T$,
 it is known \cite{Greenbaum2019} that
\begin{equation}\label{}
  \partial\mathbf{u}_{k,x,1}=\sum_{j=2}^N\frac{1}{\lambda_1-\lambda_j}\mathbf{u}_{k,x,j}
  \mathbf{u}_{k,x,j}^T\partial(\mathbf{M}_{k,x}\mathbf{M}_{k,x}^T)\mathbf{u}_{k,x,1}.
\end{equation}
where $\lambda_j$ is the $j$th eigenvalue of $\mathbf{M}_{k,x}$ corresponding to the $j$th eigenvector $\mathbf{u}_{k,x,j}$.
Here $\partial(\mathbf{M}_{k,x}\mathbf{M}_{k,x}^T)=\sum_l\mathbf{Q}_{k,l}\partial\mathbf{x}\mathbf{x}^T\mathbf{Q}_{k,l}^T +\sum_l\mathbf{Q}_{k,l}\mathbf{x}\partial\mathbf{x}^T\mathbf{Q}_{k,l}^T$. It follows that
\begin{equation}\label{eq:yTx}
  \partial\mathbf{u}_{k,x,1}=\mathbf{T}\partial\mathbf{x}
\end{equation}
where $\mathbf{T}=\mathbf{A}+\mathbf{B}$ with
\begin{equation}\label{}
  \mathbf{A} = \sum_{j=2}^N\frac{1}{\lambda_1-\lambda_j}\mathbf{u}_{k,x,j}\mathbf{u}_{k,x,j}^T
  \sum_{l=1}^N\mathbf{Q}_{k,l}
  \mathbf{x}^T
  \mathbf{Q}_{k,l}^T\mathbf{u}_{k,x,1}
\end{equation}
\begin{equation}\label{}
  \mathbf{B} = \sum_{j=2}^N \frac{1}{\lambda_1-\lambda_j}\mathbf{u}_{k,x,j}\mathbf{u}_{k,x,j}^T\sum_{l=1}^N
  \mathbf{Q}_{k,l}\mathbf{x}\mathbf{u}_{k,x,1}^T\mathbf{Q}_{k,l}.
\end{equation}
We can also write
\begin{align}\label{}
  \mathbf{T} &= \left (\sum_{j=2}^N\frac{1}{\lambda_1-\lambda_j}\mathbf{u}_{k,x,j}\mathbf{u}_{k,x,j}^T\right )
  \notag\\
  &\cdot\left (\sum_{l=1}^N\mathbf{Q}_{k,l}\left [(\mathbf{x}^T
  \mathbf{Q}_{k,l}^T\mathbf{u}_{k,x,1})\mathbf{I}_N+\mathbf{x}\mathbf{u}_{k,x,1}^T\mathbf{Q}_{k,l}
  \right]\right )
\end{align}
where the first matrix component has the rank $N-1$ and hence so does $\mathbf{T}$.

Let $\partial\mathbf{x}=\mathbf{w}$ which consists of i.i.d. elements with zero mean and variance $\sigma_w^2\ll 1$. It then follows that
\begin{equation}\label{}
  \mathcal{E}_w\{\|\partial \mathbf{u}_{k,x,1}\|^2\}=Tr\{\mathbf{T}\sigma_w^2\mathbf{T}^T\}
=\sigma_w^2\sum_{j=1}^{N-1}\sigma_j^2
\end{equation}
where $\sigma_j$ for $j=1,\cdots,N-1$ are the nonzero singular values of $\mathbf{T}$.
Since $\mathcal{E}_w\{\|\partial\mathbf{x}\|^2\}=N\sigma_w^2$, we have
\begin{equation}\label{eq:eta_k_x}
  \eta_{k,x}\doteq\sqrt{\frac{\mathcal{E}_w\{\|\partial \mathbf{u}_{k,x,1}\|^2\}}{\mathcal{E}_w\{\|\partial\mathbf{x}\|^2\}}}=
  \sqrt{\frac{1}{N}\sum_{j=1}^{N-1}\sigma_j^2}
\end{equation}
which measures a local sensitivity of $\mathbf{u}_k$  to a perturbation in $\mathbf{x}$. Since both $\mathbf{x}$ and $\mathbf{u}_k$ have the unit norm, we can view $1/\mathcal{E}_w\{\|\partial\mathbf{x}\|^2\}$ as SNR of $\mathbf{x}$ and $1/\mathcal{E}_w\{\|\partial \mathbf{u}_{k,x,1}\|^2\}$ as SNR of $\mathbf{u}_{k,x,1}$.

For each given $\mathbf{x}$, there is a small percentage of realizations of $\{\mathbf{Q}_{k,l},l=1,\cdots,N\}$ that make $\eta_{k,x}$ relatively large. To reduce $\eta_{k,x}$, we can simply prune away such bad realizations.

\begin{figure}[h]
  \centering
  \centering
  \includegraphics[width=80mm]{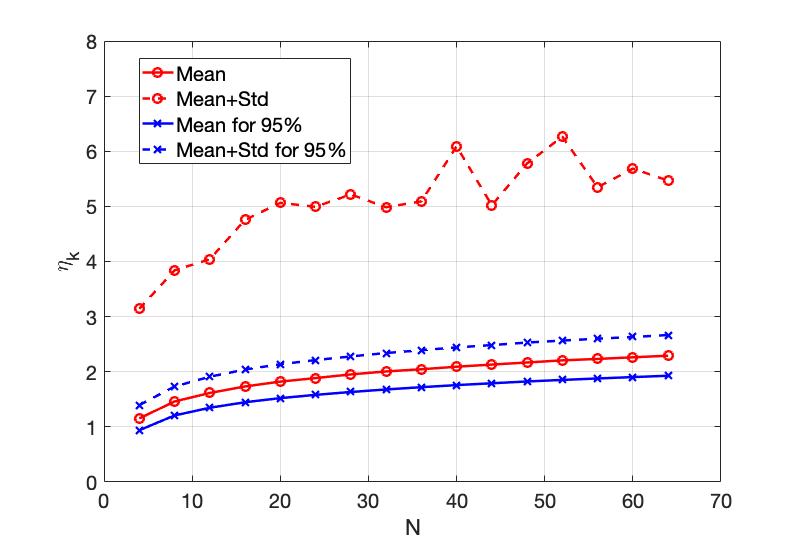}
  \caption{The mean and mean-plus-deviation of $\eta_{k,x}$ versus $N$.}
  \label{fig:etakx}
  \end{figure}

Shown in Fig. \ref{fig:etakx} are the means and means-plus-deviations of $\eta_{k,x}$ (over choices of $k$ and  $\mathbf{x}$) versus $N$, with and without pruning respectively. Here ``std'' stands for standard deviation. We see that $5\%$ pruning (or equivalently $95\%$ inclusion shown in the figure) results in a substantial reduction of $\eta_{k,x}$. We used $1000\times1000$ realizations of $\mathbf{x}$ and $\{\mathbf{Q}_{k,l},l=1,\cdots,N\}$.

  Shown in Table \ref{T:eta} are some statistics of $\eta_{k,x}$ subject to $\eta_{k,x}<2.5$. And $P_{good}$ is the probability of $\eta_{k,x}<2.5$.
\begin{table}
  \centering
  \caption{Statistics of $\eta_{k,x}$ subject to $\eta_{k,x}<2.5$ and $P_{good}$}\label{T:eta}
  \begin{tabular}{|c|c|c|c|}
     \hline
     $N$ & 16 & 32 & 64 \\
     \hline
     Mean & 1.325 & 1.489 & 1.645 \\
     \hline
     Std & 0.414 & 0.397 & 0.371 \\
     \hline
     $P_{good}$ & 0.88 & 0.84 & 0.78 \\
     \hline
   \end{tabular}
\end{table}

\subsubsection{Global relationships}

Any unit-norm vector $\mathbf{x}'$ can be written as $\mathbf{x}'=\pm \sqrt{1-\alpha}\mathbf{x}+\sqrt{\alpha}\mathbf{w}$ where $0\leq \alpha \leq 1$, and $\mathbf{w}$ is of the unit norm and satisfies $\mathbf{w}^T\mathbf{x}=0$. Then
$\|\Delta \mathbf{x}\|=\|\mathbf{x}'-\mathbf{x}\|=\sqrt{2-2\sqrt{1-\alpha}}
$.
It follows that $\|\Delta \mathbf{x}\| \leq \sqrt{2}$ and $\|\Delta\mathbf{u}_k\|\leq \sqrt{2}$.
For given $\alpha$ in $\mathbf{x}'=\pm \sqrt{1-\alpha}\mathbf{x}+\sqrt{\alpha}\mathbf{w}$, $\|\Delta \mathbf{x}\|$ is given while $\|\Delta\mathbf{u}_k\|$  still depends on $\mathbf{w}$.

\begin{figure}[h]
  \centering
  \includegraphics[width=80mm]{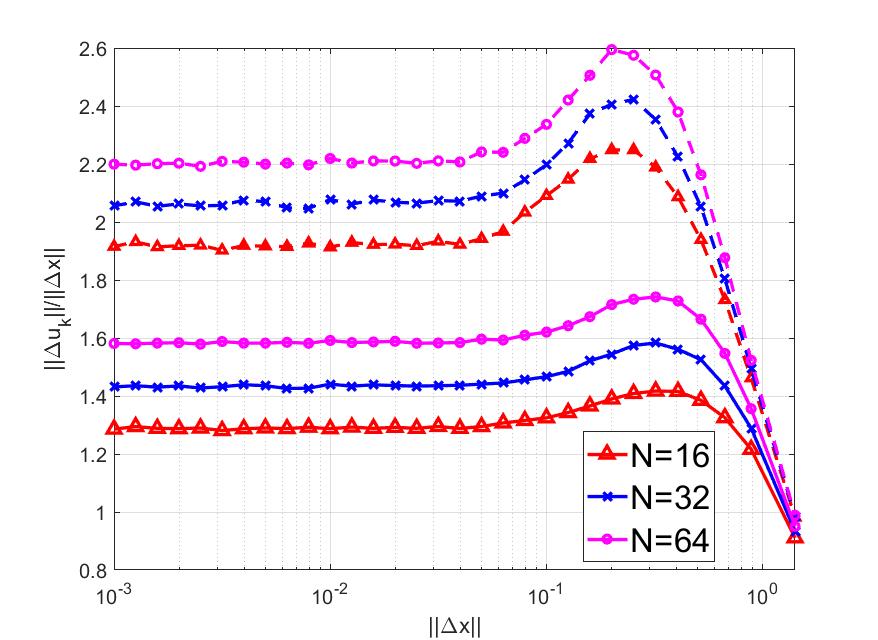}
  \caption{The means (lower three curves) and means-plus-deviations (upper three curves)  of $\frac{\|\Delta\mathbf{u}_k\|}{\|\Delta \mathbf{x}\|}$ subject to $\eta_{k,x}<2.5$.}\label{T:du_over_dx}
  \end{figure}

Shown in Fig. \ref{T:du_over_dx} are the means and means-plus-deviations of $\frac{\|\Delta\mathbf{u}_k\|}{\|\Delta \mathbf{x}\|}$ versus $\|\Delta \mathbf{x}\|$  subject to $\eta_{k,x}<2.5$. This figure is based on $1000\times1000$ realizations of $\mathbf{x}$ and $\{\mathbf{Q}_{k,l},l=1,\cdots,N\}$ under the constraint $\eta_{k,x}<2.5$.

\subsection{Correlation between Input and Output}

\subsubsection{When there is a secret key}
Recall $\mathbf{M}_{k,x}=[\mathbf{Q}_{k,1}\mathbf{x},\cdots,\mathbf{Q}_{k,N}\mathbf{x}]$. With a secret key, we can assume that $\mathbf{Q}_{k,l}$ for all $k$ and $l$ are uniformly random unitary matrices (from adversary's perspective). Then $\mathbf{u}_k$ for all $k$ and any $\mathbf{x}$ are uniformly random on $\mathcal{S}^{N-1}(1)$. It follows that $\mathcal{E}_Q\{\mathbf{u}_k\mathbf{u}_m^T\}=0$ for $k\neq m$, and $\mathcal{E}_Q\{\mathbf{u}_k\mathbf{x}^T\}=0$. Furthermore, it can be shown that $\mathcal{E}_Q\{\mathbf{u}_k\mathbf{u}_k^T\}=\frac{1}{N}\mathbf{I}_N$, i.e., the entries of $\mathbf{u}_k$ are uncorrelated with each other. Here $\mathcal{E}_Q$ denotes the expectation over the distributions of $\mathbf{Q}_{k,l}$.

\subsubsection{When there is no secret key}
In this case, $\mathbf{Q}_{k,l}$ for all $k$ and $l$ must be treated as known. But we consider typical (random but known) realizations of $\mathbf{Q}_{k,l}$ for all $k$ and $l$.

To understand the correlation between $\mathbf{x}\in \mathcal{S}^{N-1}(1)$ and $\mathbf{u}_k\in \mathcal{S}^{N-1}(1)$ subject to a fixed (but typical) set of $\mathbf{Q}_{k,l}$, we consider the following measure:
\begin{equation}\label{}
  \rho_k=N\max_{i,j}|[\mathcal{E}_x\{\mathbf{x}\mathbf{u}_k^T\}]_{i,j}|
\end{equation}
where $\mathcal{E}_x$ denotes the expectation over the distribution of $\mathbf{x}$. If $\mathbf{u}_k=\mathbf{x}$, then $\rho_k=1$. So, if the correlation between $\mathbf{x}$ and $\mathbf{u}_k$ is small, so should be $\rho_k$. For comparison, we define $\rho_k^*$ as $\rho_k$ with $\mathbf{u}_k$ replaced by a random unit-norm vector (independent of $\mathbf{x}$).

\begin{figure}[h]
  \centering
  \includegraphics[width=80mm]{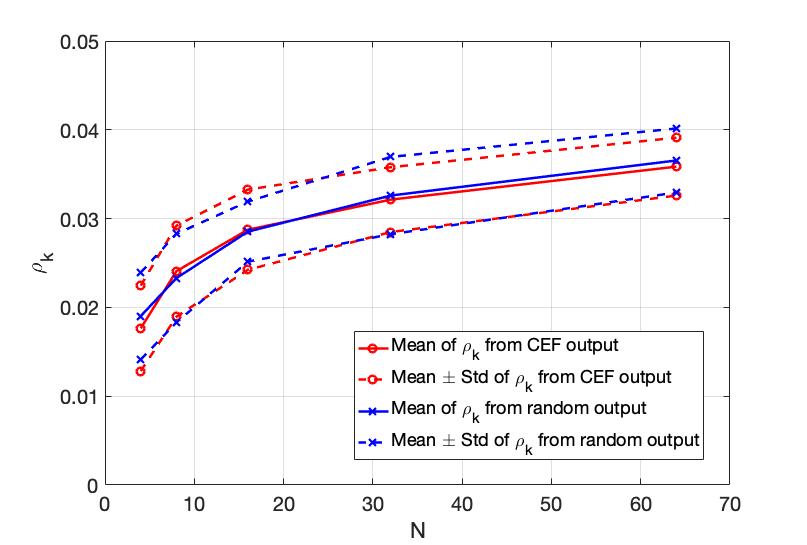}
  \caption{The means and means$\pm$deviations of $\rho_k$ (using SVD-CEF output) and $\rho_k^*$ (using random output) versus $N$ subject to $\eta_{k,x}<2.5$.}
  \label{fig:correlation}
  \end{figure}

For a different $k$, there is a different realization of $\mathbf{Q}_{k,1},\cdots,\mathbf{Q}_{k,N}$. Hence, $\rho_k$ changes with $k$.  Shown in Fig. \ref{fig:correlation} are the mean and mean$\pm$deviation of $\rho_k$ and $\rho_k^*$ versus $N$ subject to $\eta_{k,x}<2.5$. We used $10000\times100$ realizations of $\mathbf{x}$ and $\{\mathbf{Q}_{k,1},\cdots,\mathbf{Q}_{k,N}\}$. We see that $\rho_k$ and $\rho_k^*$ have virtually the same mean and deviation. (Without the constraint $\eta_{k,x}<2.5$, $\rho_k$ and $\rho_k^*$ match even better with each other.)

\subsection{Difference between Input and Output Distributions}\label{sec:entropy}

We show next that $\mathbf{u}_k$ for all $k$ have a near-zero linear correlation among themselves, and each $\mathbf{u}_k$ is nearly uniformly distributed on $\mathcal{S}^{N-1}(1)$ when $\mathbf{x}$ is uniformly distributed on $\mathcal{S}^{N-1}(1)$.

When $\mathbf{Q}_{k,1}$ for all $k$ and $l$ are independent random unitary matrices, we know that $\mathbf{u}_k$ and $\mathbf{u}_m$ for $k\neq m$ are independent of each other and $\mathcal{E}_Q(\mathbf{u}_k\mathbf{u}_m^T)=0$. Then for any typical realization of such $\mathbf{Q}_{k,1}$ for all $k$ and $l$, and for any $\mathbf{x}$, we should have $\frac{1}{K}\sum_{k=1}^K\mathbf{u}_k\mathbf{u}_{k+m}^T\approx 0$ for large $K$ and any $m\geq 1$, which means a near-zero linear correlation among $\mathbf{u}_k$ for all $k$.

\begin{figure}[h]
  \centering
  \includegraphics[width=80mm]{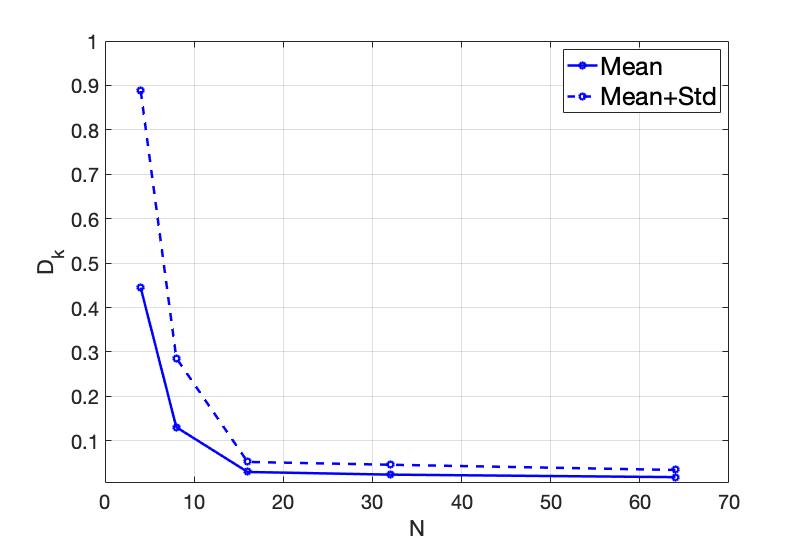}
  \caption{The mean and mean$\pm$deviation of $D_{k,v}$ versus $N$ subject to $\eta_{k,x}<2.5$.}
  \label{fig:Dkv}
  \end{figure}

To show that the distribution of $\mathbf{u}_k$ for each $k$ is also nearly uniform on $\mathcal{S}^{N-1}(1)$, we need to show that for any $k$ and any unit-norm vector $\mathbf{v}$, the probability density function (PDF) $p_{k,v}(x)$ of $\mathbf{v}^T\mathbf{u}_k$ subject to a fixed set of $\mathbf{Q}_{k,l}$ for all $l$ and random $\mathbf{x}$ on $\mathcal{S}^{N-1}(1)$ is nearly the same as the PDF $p(x)$ of any element in $\mathbf{x}$. The expression of $p(x)$ is derived in \eqref{eq:px} in Appendix \ref{sec:PDF}. The distance between $p(x)$ and $p_{k,v}(x)$ can be measured by
\begin{equation}\label{}
  D_{k,v}=\int p(x)\ln\frac{p(x)}{p_{k,v}(x)}dx\geq 0.
\end{equation}
Clearly, $D_{k,v}$ changes as $k$ and $\mathbf{v}$ change. Shown in Fig. \ref{fig:Dkv} are the mean and mean $\pm$ deviation of $D_{k,v}$ versus $N$ subject to $\eta_{k,x}<2.5$. We used $50\times1000\times 500$ realizations of $\mathbf{v}$, $\mathbf{x}$ and $\{\mathbf{Q}_{k,1},\cdots,\mathbf{Q}_{k,N}\}$. We see that $D_{k,v}$ becomes very small as $N$ increases. This means that for a large $N$, $\mathbf{u}_k$ is (at least approximately) uniformly distributed on $\mathcal{S}^{N-1}(1)$ when $\mathbf{x}$ is uniformly distributed on $\mathcal{S}^{N-1}(1)$. (Without the constraint $\eta_{k,x}<2.5$, $D_{k,v}$ versus $N$ has a similar pattern but is somewhat smaller.)

\section{Comparison between SVD-CEF and IoM-2}\label{sec:comparison}

As discussed earlier, the best known method to attack IoM-2 has the complexity $L_{N,M} 2^N$ with $L_{N,M}$ being a linear function of $M$ and $N$ respectively while the best known method to attack SVD-CEF has the complexity $P_{N,M} 2^{\zeta N}$ with $\zeta>1$ increasing with $N$ and $P_{N,M}$ being a polynomial function of $M$ and $N$. We see that SVD-CEF is much harder to attack than IoM-2 while none of the two could be shown to be easy to attack (assuming that all elements in $\mathbf{x}$ have independently random signs from the perspective of the attacker).

The complexity of forward computation of IoM-2 is less than that of SVD-CEF. The former is $\mathcal{O}(N^2)$ per (integer) sample of the output while the latter is $\mathcal{O}(N^3)$ per (real-valued) sample of the output.

To compare the noise sensitivities between SVD-CEF and IoM-2, we need to quantize the output of SVD-CEF as shown below since the output of IoM-2 is always discrete.

\subsection{Quantization of SVD-CEF}

Let the $k$th (real-valued) sample of the output of SVD-CEF at Alice due to the input vector $\mathbf{x}$ be $y_k$, and the $k$th sample of the output of SVD-CEF at Bob due to the input vector $\mathbf{x}'=\mathbf{x}+\mathbf{w}$ be $y_k'$. In the simulation, we will assume that the perturbation vector  $\mathbf{w}$ is white Gaussian, i.e., $\mathcal{N}(0,\sigma^2\mathbf{I})$.

As shown before, the PDF of $y_k$ can be approximated by \eqref{eq:px} in Appendix \ref{sec:PDF}, i.e., $f_{y_k}(y) = C_N (1-y^2)^{\frac{N-3}{2}}$ with $C_N=\frac{\Gamma(\frac{N}{2})}{\sqrt{\pi}\Gamma(\frac{N-1}{2})}$ and $-1<y<1$. To quantize $y_k$ into $n_y=\log_2N_y$ bits, Alice first over quantizes $y_k$ into $m_y=\log_2M_y$ bits with $M_y=N_yL_y$. Each of the $M_y$ quantization intervals within $(-1,1)$ is chosen to have the same probability $\frac{1}{M_y}$. For example, the left-side boundary value $t_i$ of the $i$th interval can be computed (offline) from $\int_{-1}^{t_i} f_{y_k}(y)dy = \frac{i}{M_y}$ with $i=0,1,\cdots,M_y-1$. A closed form of $\int(1-y^2)^{\frac{N-3}{2}}dy=\int \cos^{N-2}\theta d\theta $ with $y=\sin\theta$ is available for efficient bisection search of $t_i$. Specifically, $\int \cos^n \theta d\theta = \frac{\cos^{n-1}\theta \sin\theta}{n}+\frac{n-1}{n}\int \cos^{n-2}\theta d\theta$.

 The additional $l_y=\log_2L_y$ bits are used to assist the quantization of $y_k'$ at Bob. Specifically, if $y_k$ is quantized by Alice into an integer $0\leq i_k\leq M_y-1$, which has the standard binary form $b_1\cdots b_{n_y}b_{n_y+1}\cdots b_{m_y}$, then Alice keeps the first $n_y$ bits $b_1\cdots b_{n_y}$, corresponding to an integer $0\leq m_k\leq N_y-1$, and informs Bob of the last $l_y$ bits $b_{n_y+1}\cdots b_{m_y}$, corresponding to an integer $0\leq j_k\leq L_y-1$. Then the quantization of $y_k'$ by Bob is $m_k'  = arg\min_{m=0,\cdots,N_y-1} |y_k' - j_k -mL_y|$.

 If $m_k$ differs from $m_k'$, it is very likely that $m_k'=m_k\pm 1$. So, Gray binary code should be used to represent the integers $m_k$ and $m_k'$ at Alice and Bob respectively.

 We will choose $N_y=N$ in simulation. So, each of $m_k$ and $m_k'$, corresponding to each pair of $y_k$ and $y_k'$ respectively, is represented by $\log_2 N$ bits.

 The above quantization scheme is related to those for secret key generation in \cite{Wallace2010} and \cite{Ye2010}. Here, we have a virtually unlimited amount of  $y_k$ and $y_k'$ for $k\geq 1$, and a limited bit error rate after quantization is not a problem in practice such as for biometric based authentication (where ``Alice'' should be replaced by ``registration phase'' and ``Bob'' by ``validation phase'').

\subsection{Comparison of Bit Error Rates}

We can now compare the bit error rates (BERs) between the quantized SVD-CEF and the IoM-2.

To generate the $k$th output (integer) sample of IoM-2, we consider that Alice applies $N$ random permutations to  the $N\times 1$ feature vector $\mathbf{x}$ to produce $\mathbf{v}_{k,1},\cdots,\mathbf{v}_{k,N}$ respectively, and then computes the element-wise product of these vectors to produce $\mathbf{w}_k$. The index of the largest entry in $\mathbf{w}_k$ is now denoted by $0\leq m_k\leq N-1$. Bob conducts the same operations on $\mathbf{x}'=\mathbf{x}+\mathbf{w}$ to produces $0\leq m_k' \leq N-1$. We also apply Gray binary code here for IoM-2, which however has little effect on the performance.

\begin{figure}[h]
  \centering
  \includegraphics[width=80mm]{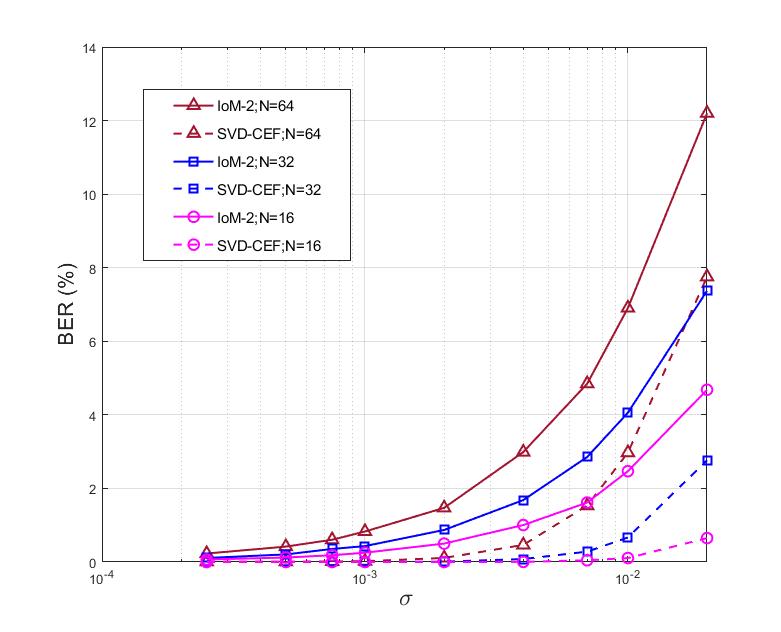}
  \caption{Bit error rates of SVD-CEF and IoM-2.}
  \label{fig:Comparison}
  \end{figure}

In Fig \ref{fig:Comparison}, we illustrate the BER performance of the quantized SVD-CEF and the IoM-2. We see that SVD-CEF consistently outperforms IoM-2. For IoM-2 and each pair of $N$ and $\sigma$, we used 1500 independent realizations of $\mathbf{x}$ according to $\mathcal{N}(0,\mathbf{I})$, and for each $\mathbf{x}$ we used an independent set of permutations and five independent realizations of $\mathbf{w}$. For SVD-CEF and each pair of $N$ and $\sigma$, we used 750 independent realizations of $\mathbf{x}$ according to $\mathcal{N}(0,\mathbf{I})$, and for each $\mathbf{x}$ we used an independent set of unitary matrices (required in $\mathbf{M}_{k,x}$ but subject to $\eta_{k,x}\leq 2.5$) and five independent realizations of $\mathbf{w}$.

Finally, it is useful to note that SVD-CEF is of type A, which is more flexible than type B. If we reduce the number $N_y$ of quantized bits per output sample, the BER of SVD-CEF can be further reduced. For IoM-2, however, if
we constrain the search among the first $L<N$ elements in $\mathbf{w}_k$, it only reduces the number of bits per output sample but does not improve the BER.

\section{Conclusion}\label{sec:conclusion}

In this paper, we have presented a development of continuous encryption functions (CEFs) that transcend the boundaries of wireless network science and biometric data science. The development of CEFs is critically important for  physical layer encryption of wireless communications and biometric template security for online Internet applications. We defined the family of CEFs to include all prior continuous ``one-way'' functions, but also expanded the scope of fundamental measures for a CEF. We showed that the dynamic random projection method and the index-of-max hashing algorithm 1 (IoM-1) are not hard to invert, the index-of-max hashing algorithm 2 (IoM-2) is not as hard to invert as it was thought to be, and the higher-order polynomials are easy to substitute. We also introduced a new family of nonlinear CEFs called SVD-CEF, which is empirically shown to be hard to attack. A statistical analysis of the SVD-CEF was provided, which reveals useful properties. The SVD-CEF is also shown to be less sensitive to noise than the IoM-2.

\appendix

\subsection{Attack of SVD-CEF via EVD Equilibrium in $\mathbf{X}$}\label{sec:attack}

We show next the details of an attack algorithm based on \eqref{eq:QXKyk}. Similar attack algorithms developed from \eqref{eq:MMyk} and \eqref{eq:MUV} are omitted. An earlier result was also reported in \cite{Hua2020b}.

It is easy to verify that  $\mathbf{X}=\alpha\mathbf{I}_N+(1-\alpha)\mathbf{x}\mathbf{x}^T$ with any  $-\infty <\alpha<\infty$ is a solution to the following
\begin{equation}\label{eq:QXKyk2}
  (\sum_{l=1}^N\mathbf{Q}_{k,l}\mathbf{X}\mathbf{Q}_{k,l}^T)\mathbf{u}_{k,x,1} = c_{k,x,1}\mathbf{u}_{k,x,1}
\end{equation}
where $c_{k,x,1}=\alpha+(1-\alpha)\sigma_{k,x,1}^2$. The expression \eqref{eq:QXKyk2} is more precise and more revealing than \eqref{eq:QXKyk} for the desired unknown matrix $\mathbf{X}$.

To ensure that $\mathbf{u}_{k,x,1}$ from \eqref{eq:QXKyk2} is unique, it is sufficient and necessary to find a $\mathbf{X}$ with the above structure and $1-\alpha\neq 0$. To ensure  $1-\alpha\neq 0$, we assume that $x_1x_2\neq 0$ where $x_1$ and $x_2$ are the first two elements of $\mathbf{x}$. Then we add  the following constraint:
\begin{equation}\label{}
  (\mathbf{X})_{1,2}=(\mathbf{X})_{2,1}=1.
\end{equation}
which is in addition to the previous condition $Tr(\mathbf{X})=1$.
Now for the expected solution structure $\mathbf{X}=\alpha\mathbf{I}_N+(1-\alpha)\mathbf{x}\mathbf{x}^T$, we have $1-\alpha=\frac{1}{x_1x_2}\neq 0$.

Note that $c_{k,x,1}$ in \eqref{eq:QXKyk2} is either the largest or the smallest eigenvalue of $\sum_{l=1}^N\mathbf{Q}_{k,l}\mathbf{X}\mathbf{Q}_{k,l}^T$ corresponding to whether $1-\alpha$ is positive or negative.


To develop the Newton's algorithm, we now take the differentiation of \eqref{eq:QXKyk2} to yield
\begin{equation}\label{}
  (\sum_{l=1}^N\mathbf{Q}_{k,l}\partial\mathbf{X}\mathbf{Q}_{k,l}^T)\mathbf{u}_k+
  (\sum_{l=1}^N\mathbf{Q}_{k,l}\mathbf{X}\mathbf{Q}_{k,l}^T)\partial\mathbf{u}_k = \partial c_k\mathbf{u}_k+c_k\partial\mathbf{u}_k
\end{equation}
where we have used $\mathbf{u}_k=\mathbf{u}_{k,x,1}$ and $c_k=c_{k,x,1}$ for convenience.
The first term is equivalent to $\mathbf{\tilde Q}_k\partial\mathbf{\tilde x}$ with
$\mathbf{\tilde Q}_k=(\sum_{l=1}^N\mathbf{u}_k^T\mathbf{Q}_{k,l}\otimes\mathbf{Q}_{k,l})$
and $\mathbf{\tilde x}=vec(\mathbf{X})$. (For basics of matrix differentiation, see \cite{Magnus}.)

Since $\mathbf{X}=\mathbf{X}^T$, there are repeated entries in $\mathbf{\tilde x}$. We can write $\mathbf{\tilde x}=[\mathbf{\tilde x}_1^T,\cdots,\mathbf{\tilde x}_N^T]^T$ with $\mathbf{\tilde x}_n=[\tilde x_{n,1},\cdots,\tilde x_{n,N}]^T$ and $\tilde x_{i,j}=\tilde x_{j,i}$ for all $i\neq j$. Let $\mathbf{\hat x}$ be the vectorized form of the lower triangular part of $\mathbf{X}$. Then it follows that
\begin{equation}\label{}
  \mathbf{\tilde Q}_k\partial\mathbf{\tilde x} = \mathbf{\hat Q}_k\partial\mathbf{\hat x}
\end{equation}
where $\mathbf{\hat Q}_k$ is a compressed form of $\mathbf{\tilde Q}_k$ as follows. Let $\mathbf{\tilde Q}_k=[\mathbf{\tilde Q}_{k,1},\cdots,\mathbf{\tilde Q}_{k,N}]$ with $\mathbf{\tilde Q}_{k,n}=[\mathbf{\tilde q}_{k,n,1},\cdots,\mathbf{\tilde q}_{k,n,N}]$. For all $1\leq i<j\leq N$, replace $\mathbf{\tilde q}_{k,i,j}$ by $\mathbf{\tilde q}_{k,i,j}+\mathbf{\tilde q}_{k,j,i}$, and then drop $\mathbf{\tilde q}_{k,j,i}$. The resulting matrix is $\mathbf{\hat Q}_k$.

The differential of $Tr(\mathbf{X})=1$ is $Tr(\partial\mathbf{X})=0$ or equivalently $\mathbf{t}^T\partial\mathbf{\hat x}=0$ where $\mathbf{t}^T=[\mathbf{t}_1^T,\cdots,\mathbf{t}_N^T]$ and $\mathbf{t}_n^T=[1,\mathbf{0}_{1\times (N-n)}]^T$.


Combining the above for all $k$ along with $\mathbf{u}_k^T\partial\mathbf{u}_k=0$ (due to the norm constraint $\|\mathbf{u}_k\|^2=1$) for all $k$, we have
\begin{equation}\label{eq:AxAyAz}
  \mathbf{A}_x\partial\mathbf{\hat x}+\mathbf{A}_u\partial\mathbf{u}+\mathbf{A}_z\partial\mathbf{z}=0
\end{equation}
where
\begin{equation}\label{}
  \mathbf{A}_x=\left [ \begin{array}{c}
  \mathbf{t}^T\\
                         \mathbf{\hat Q}_1 \\
                         \cdots\\
                         \mathbf{\hat Q}_K \\
                         \mathbf{0}_{K\times \frac{1}{2}N(N+1)}
                       \end{array}
  \right ]
\end{equation}
\begin{equation}\label{}
  \mathbf{A}_u=\left [\begin{array}{c}
  \mathbf{0}_{1\times NK}\\
                        diag(\mathbf{G}_{1,x},
\cdots,\mathbf{G}_{K,x}) \\
                        diag(\mathbf{u}_1^T,\cdots,\mathbf{u}_K^T)
                      \end{array}
   \right ],
\end{equation}
\begin{equation}\label{}
  \mathbf{A}_z=\left [\begin{array}{c}
 \mathbf{0}_{1\times K} \\
  -diag(\mathbf{u}_1,\cdots,\mathbf{u}_K)\\
  \mathbf{0}_{K\times K}
    \end{array}
 \right ]
\end{equation}
with $\mathbf{G}_{k,x}=\mathbf{M}_{k,x}\mathbf{M}_{k,x}^T-c_k\mathbf{I}_M$.

Now we partition $\mathbf{u}$ into two parts: $\mathbf{u}_a$ (known) and $\mathbf{u}_b$ (unknown). Also partition $\mathbf{A}_u$ into $\mathbf{A}_{u,a}$ and $\mathbf{A}_{u,b}$ such that
$\mathbf{A}_u\partial\mathbf{u} =\mathbf{A}_{u,a}\partial\mathbf{u}_a+\mathbf{A}_{u,b}\partial\mathbf{u}_b$. Since $(\mathbf{X})_{1,2}=(\mathbf{X})_{2,1}=1$, we also let $\mathbf{\hat x}_0$ be $\mathbf{\hat x}$ with its second element removed, and $\mathbf{A}_{x,0}$ be $\mathbf{A}_x$ with its second column removed. It follows from \eqref{eq:AxAyAz} that
\begin{equation}\label{eq:AaBb2}
  \mathbf{A}\partial\mathbf{a}+\mathbf{B}\partial\mathbf{b}=0
\end{equation}
where $\mathbf{a}=\mathbf{u}_a$, $\mathbf{b}=[\mathbf{\hat x}_0^T, \mathbf{u}_b^T,\mathbf{z}^T]^T$, $\mathbf{A}=\mathbf{A}_{u,a}$,
$\mathbf{B}=[\mathbf{A}_{x,0},\mathbf{A}_{u,b},\mathbf{A}_z]$.

Based on \eqref{eq:AaBb2}, the Newton's algorithm is
\begin{equation}\label{eq:Netwon2}
  \left [ \begin{array}{c}
            \mathbf{\hat x}_0^{(i+1)} \\
            *
          \end{array}
  \right ]=\left [ \begin{array}{c}
            \mathbf{\hat x}_0^{(i)} \\
            *
          \end{array}
  \right ]-\eta(\mathbf{B}^T\mathbf{B})^{-1}\mathbf{B}^T\mathbf{A}(\mathbf{u}_a
  -\mathbf{u}_a^{(i)})
\end{equation}
where the terms associated with $*$ are not needed, $\mathbf{u}_a^{(i)}$ is the $i$th-step ``estimate'' of the known vector $\mathbf{u}_a$ (through forward computation) based on the $i$-step estimate $\mathbf{\hat x}_0^{(i)}$ of the unknown vector $\mathbf{\hat x}_0$.
This algorithm requires $NyK\geq \frac{1}{2}N(N+1)-1$ in order for $\mathbf{B}$ to have full column rank.

For a random initialization around $\mathbf{X}$, we can let $\mathbf{X}'=(1-\beta)\mathbf{X} +\beta \mathbf{W}$ where $\mathbf{W}$ is a symmetric random matrix with $Tr(\mathbf{W})=1$. Furthermore, $(\mathbf{W})_{1,2}=(\mathbf{W})_{2,1}$ is such that $(\mathbf{X}')_{1,2}=(\mathbf{X}')_{2,1}=1$. Keep in mind that at every step of iteration, we keep $(\mathbf{X}^{(i)})_{1,2}=(\mathbf{X}^{(i)})_{2,1}=1$.

Upon convergence of $\mathbf{X}$, we can also update $\mathbf{x}$ as follows. Let the eigenvalue decomposition of $\mathbf{X}$ be $\mathbf{X}=\sum_{i=1}^N\lambda_i\mathbf{e}_i\mathbf{e}_i^T$ where $\lambda_1>\lambda_2>\cdots>\lambda_N$. Then the update of $\mathbf{x}$ is given by $\mathbf{e}_1$ if $1-\alpha>0$ or by $\mathbf{e}_N$ if $1-\alpha<0$. With each renewed $\mathbf{x}$, there are a renewed $\alpha$ and hence a renewed $\mathbf{X}$ (i.e., by setting $\mathbf{X}=\alpha\mathbf{I}+(1-\alpha)\mathbf{x}\mathbf{x}^T$ with $1-\alpha=\frac{1}{x_1x_2}$). Using the new $\mathbf{X}$ as the initialization, we can continue the search using \eqref{eq:Netwon2}.

The performance of the algorithm \eqref{eq:Netwon2} is discussed in section \ref{sec:performance_of_attack}.

\subsection{Distributions of Elements of a Uniformly Random Vector on Sphere}\label{sec:PDF}
Let $\mathbf{x}$ be uniformly random on $\mathcal{S}^{n-1}(r)$. This vector can be parameterized as follows:
\begin{align}\label{}
  x_1&=r\cos\theta_1\notag\\
  x_2&=r\sin\theta_1\cos\theta_2\notag\\
  \cdots\notag\\
  x_{n-1}&=r\sin\theta_1\cdots\sin\theta_{n-2}\cos\theta_{n-1}\notag\\
  x_n&=r\sin\theta_1\cdots\sin\theta_{n-2}\sin\theta_{n-1}\notag
\end{align}
where $0<\theta_i\leq \pi$ for $i=1,\cdots, n-2$, and $0<\theta_{n-1}\leq 2\pi$.
According to Theorem 2.1.3 in \cite{Muirehead1982}, the differential of the surface area on $\mathcal{S}^{n-1}(r)$ is
\begin{equation}\label{}
  dS^{n-1}(r) = r^{n-1}\sin^{n-2}\theta_1\sin^{n-3}\theta_2\cdots\sin\theta_{n-2}d\theta_1\cdots d\theta_{n-1}
\end{equation}
We know that $\int_{\mathcal{S}^{n-1}(r)}dS^{n-1}(r)=|\mathcal{S}^{n-1}(r)|=
\frac{2\pi^{n/2}}{\Gamma(\frac{n}{2})}r^{n-1}$.
Hence,
the PDF of $\mathbf{x}$ is
\begin{equation}\label{}
  f_x(\mathbf{x})=\frac{1}{|\mathcal{S}^{n-1}(r)|}.
\end{equation}

\subsubsection{Distribution of one element in $\mathbf{x}$}
We can rewrite $\int_{\mathcal{S}^{n-1}(r)}f_x(\mathbf{x})dS^{n-1}(r)=1$ as
\begin{equation}\label{}
  \int_{\theta_1}\left [\int_{\mathcal{S}^{n-2}(r\sin\theta_1)}f_x(\mathbf{x})rdS^{n-2}(r\sin\theta_1)\right ]d\theta_1=1
\end{equation}
or equivalently
\begin{equation}\label{}
  \int_{\theta_1}\left [\frac{|\mathcal{S}^{n-2}(r\sin\theta_1)|}{|\mathcal{S}^{n-1}(r)|}r\right ]d\theta_1=1.
\end{equation}
Hence the PDF of $\theta_1$ is
\begin{equation}\label{}
  f_{\theta_1}(\theta_1)=\frac{|\mathcal{S}^{n-2}(r\sin\theta_1)|}{|\mathcal{S}^{n-1}(r)|}r.
\end{equation}
To find the PDF of $x_1=r\cos\theta_1$, we have
\begin{equation}\label{}
  f_{x_1}(x_1)=f_{\theta,1}(\theta_1)\frac{1}{\left |\frac{dx_1}{d\theta_1}\right |}=\frac{f_{\theta,1}(\theta_1)}{|r\sin\theta_1|}
\end{equation}
where $r\sin\theta_1=\sqrt{r^2-x_1^2}$. Therefore, combining all the previous results yields
\begin{equation}\label{}
  f_{x_1}(x_1)=\frac{\Gamma(\frac{n}{2})}{\sqrt{\pi}\Gamma(\frac{n-1}{2})}
  \frac{(r^2-x_1^2)^{\frac{n-3}{2}}}{r^{n-2}}
\end{equation}
where $-r\leq x_1\leq r$.

If $r=1$, we have
\begin{equation}\label{eq:px}
  f_{x_1}(x_1)=\frac{\Gamma(\frac{n}{2})}{\sqrt{\pi}\Gamma(\frac{n-1}{2})}
  (1-x_1^2)^{\frac{n-3}{2}}
\end{equation}
where $-1\leq x_1\leq 1$. This is the PDF $p(x)$ in section \ref{sec:entropy}.

Due to symmetry, we know that $x_i$ for any $i$ has the same PDF as $x_1$. Also note that if $n=3$, $f_{x_1}(x)$ is a uniform distribution.

\end{document}